\newif\ifTimes\newif\ifJSS\JSStrue
\JSSfalse%<- comment this (in and out) - PLUS set JSS <- ... in 01_setup_fcts_lapply.Rnw
\ifJSS\Timesfalse
\documentclass[article]{jss}
\else\Timestrue
\documentclass[article,nojss]{jss}
\fi
\expandafter\let\csname Schunk\endcsname\relax
\expandafter\let\csname endSchunk\endcsname\relax

\author{Marius\ Hofert\thanks{The\ author\ (Willis\ Research\ Fellow)\ thanks\
    Willis\ Re\ for\ financial\ support\ while\ this\ work\ was\ being\
    completed.}\\ ETH\ Zurich \And Martin M\"achler\\ ETH\ Zurich}
\Address{
  Marius Hofert\\
  RiskLab, Department of Mathematics\\
  ETH Zurich\\
  8092 Zurich, Switzerland\\
  E-mail: \email{marius.hofert@math.ethz.ch}\\
  URL: \url{http://www.math.ethz.ch/~hofertj/}

  \bigskip
  Martin M\"achler\\
  Seminar f\"ur Statistik, HG G~16\\
  ETH Zurich\\
  8092 Zurich, Switzerland\\
  E-mail: \email{maechler@stat.math.ethz.ch}\\
  URL: \url{http://stat.ethz.ch/people/maechler}
}
\title{Parallel and other simulations in \R\ made easy:\\ An end-to-end study}
\Plainauthor{Marius Hofert, Martin M{\"a}chler}% note: 'M\"achler' does not work with SweavePDF!; MH: yes it does :-)
\Plaintitle{Parallel and other simulations in R made easy: An end-to-end study} % without formatting

\Abstract{\hspace{-1.84em} It is shown how to set up, conduct, and analyze large simulation
  studies with the new \R\ package
  \begin{align*}
    \text{\pkg{simsalapar}} = \text{\pkg{sim}ulations \pkg{s}implified \pkg{a}nd \pkg{la}unched
        \pkg{par}allel}.% or: simulations simplified a lot via? after? as? parallel(ization); a = activity,
  \end{align*}
  A simulation study typically starts with determining a collection of input
  variables and their values on which the study depends, such as sample sizes,
  dimensions, types and degrees of dependence, estimation methods,
  etc. Computations are desired for all combinations of these variables. If
  conducting these computations sequentially is too time-consuming, parallel
  computing can be applied over all combinations of select variables. The final result
  object of a simulation study is typically an array. From this array, summary
  statistics can be derived and presented in terms of (flat contingency or
  \LaTeX) tables or visualized in terms of (matrix-like) figures.

  The \R\ package \pkg{simsalapar} provides several tools to achieve the above
  tasks. Warnings and errors are dealt with correctly, various
  seeding methods are available, and run time
  is measured. Furthermore, tools for analyzing the results via tables or
  graphics are provided. In contrast to
  rather minimal examples typically found in \R\ packages or vignettes, an
  end-to-end, not-so-minimal simulation problem from the realm of quantitative
  risk management is given. The concepts presented and solutions provided by
  \pkg{simsalapar} may be of interest to students, researchers, and
  practitioners as a how-to for conducting realistic, large-scale simulation
  studies in \R. Also, the development of the package revealed useful
  improvements to \R\ itself, which are available in \R\ 3.0.0.
}

\Keywords{\R, simulation, parallel computing, data analysis}
\Plainkeywords{R, simulation, parallel computing, data analysis}

\usepackage[T1]{fontenc}% for correct hyphenation and T1 encoding
\usepackage{lmodern}% latin modern font
\usepackage[american]{babel}% for American English
\usepackage{microtype}% for character protrusion and font expansion (only with pdflatex)

\usepackage{eso-pic}% for setting up background
\usepackage{rotating}% for landscape tables
\usepackage{amsmath}% sophisticated mathematical formulas with amstex (includes \text{})
\usepackage{mathtools}% fix amsmath deficiencies
\usepackage{amssymb}% sophisticated mathematical symbols with amstex
\usepackage{amsthm}% theorem environments
\usepackage{etoolbox}% for preto
\usepackage{bm}% for bold math symbols
\usepackage{bbm}% only for indicator functions
\usepackage{enumitem}% sophisticated itemize, enumerate
\usepackage{graphicx}% for including figures
\usepackage{grffile}% extending the file name processing of graphicx
\usepackage{tikz}% sophisticated graphics package
\usetikzlibrary{decorations.pathmorphing, decorations.text}% for text decorations
\usepackage{wrapfig}% for wrapping figures and tables
\usepackage{tabularx}% for tables
\usepackage{dcolumn}% for special column types and alignment requirements
\usepackage{siunitx}% for typesetting numbers with units; also for alignment in tables
\usepackage{booktabs}% for table layout
\usepackage{multirow}% for table entries spanning several row
\usepackage{Sweave}% for working with R
\usepackage{listings}% for including source code

\usepackage{catchfile}

\newcommand{\getvar}[2]{%
  \CatchFileEdef#1{"|kpsewhich -var-value #2"}{\endlinechar=-1 }%
}

\getvar{\username}{USER}

\newif\ifMM
\def\MM{maechler}
\ifx\MM\username \MMtrue \else \MMfalse \fi

\ifMM\else
\usepackage{vruler}% for showing line numbers
\fi

\xdefinecolor{lightgray}{RGB}{247, 247, 247}% R's col2rgb("gray97"); #F7F7F7
\xdefinecolor{semilightgray}{RGB}{240, 240, 240}% R's col2rgb("gray94"); #F0F0F0
\xdefinecolor{middlegray}{RGB}{127, 127, 127}% R's col2rgb("gray50"); #7F7F7F
\xdefinecolor{seashell}{RGB}{255, 245, 238}% R's col2rgb("seashell"); "#FFF5EE"
\xdefinecolor{yellow}{RGB}{255, 255, 128}% R's self-defined (weakened yellow); #FFFF80
\xdefinecolor{orange}{RGB}{238, 118, 0}% R's col2rgb("darkorange2"); #EE7600
\xdefinecolor{red}{RGB}{178, 34, 34}% R's col2rgb("firebrick"); #B22222
\xdefinecolor{blue}{RGB}{58, 95, 205}% R's col2rgb("royalblue3"); #3A5FCD
\xdefinecolor{deepskyblue}{RGB}{0, 154, 205}% R's col2rgb("deepskyblue3"); #009ACD
\xdefinecolor{chocolate}{RGB}{205, 102, 29}% R's col2rgb("chocolate3"); #CD661D

\setlist{% general list settings (enumitem's itemize, enumerate, and description)
  align=left,% left-aligned enumerate
  labelsep=*,% align all item bodies vertically
  leftmargin=*,% begin item content at a variable place depending on the item; use \parindent to set it exactly to \parindent
  topsep=1mm,% space before enumerate
  itemsep=0mm% space between enumerate items
}
\setlist[itemize,1]{label={\protect\rule[0.18em]{0.36em}{0.36em}\ }}% itemize label on level 1
\setlist[itemize,2]{label={\protect\raisebox{0.12em}{\resizebox{0.48em}{0.48em}{$\blacktriangleright$}}\ }}% itemize label on level 2
\setlist[itemize,3]{label={\protect\rule[0.32em]{0.62em}{0.08em}\ }}% itemize label on level 3
\setlist[enumerate,1]{label=\arabic*)}% enumerate label on level 1
\setlist[enumerate,2]{label=\arabic{enumi}.\arabic*)}% enumerate label on level 2
\setlist[enumerate,3]{label=\arabic{enumi}.\arabic{enumii}.\arabic*)}% enumerate label on level 3

\newenvironment{Schunk}{\vspace{0.25\baselineskip}}{\vspace{0.25\baselineskip}}% adjust space before/after chunk

\lstdefinestyle{input}{
  backgroundcolor=\color{semilightgray},% background color
  commentstyle=\itshape\color{chocolate},% comment style
  keywordstyle=\color{blue},% keyword style
  stringstyle=\color{deepskyblue},% string style
  numbers=left,% display line numbers on the left side
  numberstyle=\color{middlegray}\tiny% use small line numbers
}
\lstdefinestyle{output}{
  backgroundcolor=\color{lightgray}% background color
}

\lstdefinestyle{Lstyle}{
  language=[LaTeX]TeX,% set programming language
  texcs={},% texcs
  otherkeywords={}% undefine otherkeywords
}

\lstnewenvironment{Linput}[1][]{%
  \lstset{style=input, style=Lstyle}
  #1% content
}{\vspace{-0.25\baselineskip}}% note: -\baselineskip leads to no space

\lstnewenvironment{Loutput}[1][]{%
  \lstset{style=output, style=Lstyle}
  #1% content
}{\vspace{-0.25\baselineskip}}% note: -\baselineskip leads to no space

\lstdefinestyle{Rstyle}{
  language=R,% set programming language
  literate={<-}{{$\bm\leftarrow$}}2{<<-}{{$\bm{\mathrel{\bm\leftarrow\mkern-14mu\leftarrow}$}}}2{<=}{{\raisebox{0.6pt}{\scalebox{0.8}{$\bm\le$}}}}2{>=}{{\raisebox{0.6pt}{\scalebox{0.8}{$\bm\ge$}}}}2{!=}{{$\bm\neq$}}2,% item to replace, text, length of chars
  keywords={if, else, repeat, while, function, for, in, next, break},% keywords; see R language manual, /usr/local/texlive/2012/texmf-dist/tex/latex/listings/lstlang3.sty
  otherkeywords={}% undefine otherkeywords to remove !,!=,~,$,*,\&,\%/\%,\%*\%,\%\%,<-,<<-,_,/
}

\expandafter\let\csname Sinput\endcsname\relax
\expandafter\let\csname endSinput\endcsname\relax
\expandafter\let\csname Soutput\endcsname\relax
\expandafter\let\csname endSoutput\endcsname\relax

\lstnewenvironment{Sinput}[1][]{%
  \lstset{style=input, style=Rstyle}
  #1% content
}{\vspace{-0.25\baselineskip}}% note: -\baselineskip leads to no space

\lstnewenvironment{Soutput}[1][]{%
  \lstset{style=output, style=Rstyle}
  #1% content
}{\vspace{-0.25\baselineskip}}% note: -\baselineskip leads to no space

\newcommand*{\R}{\textsf{R}}
\newcommand*{\LL}{\lstinline[{language=[LaTeX]TeX}, basicstyle=\ttfamily]}% for inline LaTeX code
\newcommand*{\cmd}{\lstinline[basicstyle=\ttfamily]}% for shell commands, file names etc.
\newcommand*{\file}[1]{\texttt{#1}}% for file names or endings
\renewcommand*{\cite}[2][]{\citet[#1]{#2}}

\newif\ifstarttheorem
\newtheoremstyle{mythmstyle}%
{0.5em}% space above
{0.5em}% space below
{}% body font
{}% indent amount
{\sffamily\bfseries\global\starttheoremtrue}% head font
{}% punctuation after head
{\newline}% space after head
{\thmname{#1}\ \thmnumber{#2}\ \thmnote{(#3)}}% head spec
\theoremstyle{mythmstyle}% activate style
\newtheorem{definition}{Definition}[section]

\newtheorem{remark}[definition]{Remark}

\renewcommand*\proofname{Proof}


\makeatletter
\preto\itemize{%
  \if@inlabel
    \ifstarttheorem
      \mbox{}\par\nobreak\vskip\glueexpr-\parskip-\baselineskip+0.3em\relax\hrule\@height\z@
      \global\starttheoremfalse%
    \fi%
  \fi%
 \def\tempa{proof}%
 \ifx\tempa\mycurrenvir
    \ifstarttheorem
      \mbox{}\par\nobreak\vskip\glueexpr-\parskip-\baselineskip+0.3em\relax\hrule\@height\z@
      \global\starttheoremfalse%
    \fi%
 \fi%
}
\preto\enditemize{\global\starttheoremfalse}
\makeatother

\makeatletter
\preto\enumerate{%
  \if@inlabel
    \ifstarttheorem
      \mbox{}\par\nobreak\vskip\glueexpr-\parskip-\baselineskip+0.3em\relax\hrule\@height\z@
      \global\starttheoremfalse%
    \fi%
  \fi%
 \def\tempa{proof}%
 \ifx\tempa\mycurrenvir
    \ifstarttheorem
      \mbox{}\par\nobreak\vskip\glueexpr-\parskip-\baselineskip+0.3em\relax\hrule\@height\z@
      \global\starttheoremfalse%
    \fi%
 \fi%
}
\preto\endenumerate{\global\starttheoremfalse}
\makeatother

\newcommand*{\IR}{\mathbbm{R}}

\newcommand*{\VaR}{\operatorname{VaR}}

\begin{document}
\ifMM\else% MM likes "clean paper" (and wants to have people *read* it)
% MH: Yes, but the ruler is typically good for feedback and the watermark is
%     is required such that a journal can't say "... it has already been out..."
% watermark
% \AddToShipoutPicture{% set up watermark on every page
%   \begin{tikzpicture}[remember picture, overlay]
%     \node[scale=9,rotate=54.74,color=black!18] at (current
%     page.center){\normalfont\sffamily Draft};
%   \end{tikzpicture}%
% }
% ruler
\setvruler[10pt][1][1][4][1][0pt][0pt][0pt][\textheight]%
\fi

\section{Introduction}
Realistic mathematical or statistical models are often complex and not analytically
tractable, thus require to be evaluated by simulation. In many areas such as
finance, insurance, or statistics, it is therefore necessary to set up, conduct,
and analyze simulation studies. Apart from minimal examples which address
particular tasks, one often faces more difficult setups with a complex
simulation problem at hand. For example, if a comparably small simulation
already reveals an interesting result, it is often desired to conduct a larger
study, involving more parameters, a larger sample size, or more simulation
replications. However, run time for sequentially computing results for all
variable combinations may now be too large. It may thus be beneficial to apply parallel
computing for select variable combinations, be it on a multi-core processor with
several central processing units (\emph{cores}), or on a network (\emph{cluster}) with
several computers (\emph{nodes}). This adds another level of
difficulty to solving the initial task. Users such as students (for a master or
Ph.D.\ thesis, for example), researchers (for investigating the performance of a
new statistical model), or practitioners (for computing model outputs in a short
amount of time or validating internal models), are typically not primarily interested in the technical details
of parallel computing, especially when it comes to more involved tasks such as
correctly advancing a random number generator stream to guarantee
reproducibility while having different seeds on different nodes. Furthermore,
numerical issues often distort simulation results but remain undetected,
especially if they happen rarely or are not captured correctly. These issues are
either not, or not sufficiently addressed in examples, vignettes, or other
packages one would consult when setting up a simulation study.

In this paper, we introduce and present the new \R\ package \pkg{simsalapar} and
show how it can be used to set up, conduct, and analyze a simulation study in \R. It
extends the functionality of several other \R\ packages\footnote{For example,
\pkg{simSummary}, \pkg{ezsim}, \pkg{harvestr}, and \pkg{simFrame}.}. In our
view, a simulation
study typically consists of the following parts:
\begin{enumerate}
\item\label{Step1} \emph{Setup}: The scientific problem; how to translate it to
  a setup of a simulation study; breaking down the problem into different layers
  and implementing the main, problem-specific function. These tasks are addressed
  in Sections \ref{sec:translate}--\ref{sec:lapply} after introducing our
  working example in the realm of quantitative risk management in Section
  \ref{sec:scientific:q}.
\item \emph{Conducting the simulation}: Here, approaches of how to compute in parallel
  with \R are presented. They depend on whether the simulation study is run on one machine
  (node) with a multi-core processor or on a cluster with several nodes. This is
  addressed in Section \ref{sec:parallel}.
\item\label{Step3} \emph{Analyzing the results}: How results of a simulation
  study can be presented with tables or graphics. This is done in
  Section \ref{sec:analysis}.
\end{enumerate}
In Section~\ref{sec:behind} we show additional and more advanced
computations which are not necessary for understanding the paper. They rather
emphasize what is going on ``behind the scenes'' of \pkg{simsalapar}, provide further
functionality, explanations of our ansatz, and additional checks
conducted. Section~\ref{sec:conclusion} concludes.
%% \pkg{simSummary}:  one function simSummary(), get list of "inner" and
%%                    "outer" parameters / results.
%%                  Produces "table" like R objects..
%% Vignette design:  n.sim, one "grid" or "inner", many "other"
%%
%% \pkg{ezsim}:  ezsim() -> object with quite flexible summary() and plot() methods
%%            ++ : much more graphics than tables
%%            ++ : "varList"-like object ("banker" (instead of "other")) ;
%%	instead of doOne() have
%%      {dgp(): data-gener-proc , estimator() - each user specified function}
%%
%% \pkg{harvestr}: Didactically nice idea {generality?};
%%  ++ Basic idea (as our): What harvestr brings to the picture is
%%     >> abstractions of the process of performing simulation <<
%%  ++ emphasis on correct seeding for parallel
%%  ++ timing  {and caching (correctly ??) via \pkg{digest}}
%%   some small examples, one with lme4.
%%
%% \pkg{simFrame}: Rcpp based, S4 classes and methods (many!)
%% --> JSS paper 2010  http://www.jstatsoft.org/v37/i03/
%% 	An Object-Oriented Framework for Statistical
%% 	Simulation: The R Package simFrame
%% ++ save seeds (how? / ..)
%% plot() method for sim.result
%%  "longer" examples // robustness: contamination + NA's
%%  "design-based simulation" for survey sampling ..

As a working example throughout the paper, we present a simulation problem from
the realm of quantitative risk management. The example is minimal in the sense
that it can still be run on a standard computer and does not require access to a
cluster. However, it is not too minimal in that it covers a wide range of possible
problems a simulation study might face. We believe this to be useful for users
like students, researchers, and practitioners, who often need, or would like,
to implement simulation studies of similar kind, but miss guidance and an
accompanying package of how this can be achieved.

% patchDVI setup
\Sconcordance{concordance:01_setup_fcts_lapply.tex:01_setup_fcts_lapply.Rnw:%
1 8 1 1 6 90 1 1 24 4 1 1 3 17 0 1 2 111 1 1 -142 1 0 1 20 19 %
0 1 9 4 0 1 119 6 1 1 2 10 0 1 1 7 0 1 2 2 1 1 2 1 0 1 1 10 0 %
1 4 45 1 1 36 38 0 1 2 1 1 1 24 146 1 1 4 6 0 1 13 2 1 1 2 4 0 %
1 5 28 0 1 2 12 0 1 9 4 1}

% Sweave options

 % use pdfcrop to crop .pdf
 % put figures in ./ and name them with prefix "fig-"
%\VignetteDepends{simsalapar}
%\VignetteDepends{copula}
% Sweave options from within R

\section{How to set up and conduct a simulation study}\label{sec:setup}
\subsection{The scientific problem}\label{sec:scientific:q}
As a simulation problem, we consider the task of estimating quantiles of a
distribution function of the sum of dependent random variables. This is a
statistical problem from the realm of quantitative risk management, where the
distribution function under consideration is that of losses, which, for example,
a bank faces when customers default and are unable to repay
their loans. The corresponding quantile function is termed
\emph{Value-at-Risk}. According to the Basel II/III rules of banking
supervision, banks have to compute Value-at-Risk at certain (high) quantiles as
a measure of risk they face and money they have to put aside to account for such
losses and to avoid bankruptcy.

In the language of mathematics, this can be made precise as follows. Let $S_{t,j}$ denote the
value of the $j$th of $d$ stocks at time $t\ge0$. The value of a
portfolio with these $d$ stocks at time $t$ is thus
\begin{align*}
  V_t=\sum_{j=1}^d\beta_j S_{t,j},
\end{align*}
where $\beta_1,\dots,\beta_d$ denote weights, typically the number of
shares of stock $j$ in the portfolio. Considering the
logarithmic stock prices as \emph{risk factors}, the \emph{risk-factor changes} are
given by
\begin{align}\label{eq:X_t}
  X_{t+1,j}=\log(S_{t+1,j})-\log(S_{t,j})=\log(S_{t+1,j}/S_{t,j}),\quad j\in\{1,\dots,d\}.
\end{align}
Assume that all quantities at time point $t$ (interpreted as today) are known,
and we are interested in the time point $t+1$ (one period ahead, for example one year). The
\emph{loss} of the portfolio at $t+1$ can therefore be expressed as
\begin{align}\label{eq:L_t}
  L_{t+1}&=-(V_{t+1}-V_t)=-\sum_{j=1}^d \beta_j (S_{t+1,j}-S_{t,j}) =
                        -\sum_{j=1}^d \beta_j S_{t,j}(\exp(X_{t+1,j})-1),\\\nonumber
                        &=-\sum_{j=1}^d  w_{t,j}       (\exp(X_{t+1,j})-1)
\end{align}
that is, in terms of the known weights $w_{t,j}$ (at time $t$, $\beta_j$ and
$S_{t,j}$, $j\in\{1,\dots,d\}$, are
known), and the unknown risk-factor
changes. \emph{Value-at-Risk} ($\VaR_\alpha$) of $L_{t+1}$ at
\emph{level} $\alpha\in(0,1)$ is given by
\begin{align}\label{eq:VaR.L}
  \VaR_{\alpha}(L_{t+1})=F^-_{L_{t+1}}(\alpha),
\end{align}
where $F^-_{L_{t+1}}(y)=\inf\{x\in\IR:F_{L_{t+1}}(x)\ge y\}$ denotes the quantile function of the
distribution function $F_{L_{t+1}}$ of $L_{t+1}$ (equal to the ordinary inverse
$F^{-1}_{L_{t+1}}$ if $F_{L_{t+1}}$ is continuous and strictly increasing;
see \cite{embrechtshofert2013c} for more details about such functions).

For simplicity, we drop the time index $t+1$ in what follows. Let $\bm{X}=(X_1,\dots,X_d)$ be
the $d$-dimensional vector of (possibly) dependent risk-factor changes. By
\cite{sklar1959}, its distribution function $H$ can be expressed as
\begin{align*}
  H(\bm{x})=C(F_1(x_1),\dots,F_d(x_d)),\quad\bm{x}\in\IR^d,
\end{align*}
for a copula $C$ and the marginal distribution functions $F_1,\dots,F_d$ of
$H$. A \emph{copula} is a distribution function with standard uniform univariate
margins; for an introduction to copulas, see \cite{nelsen2006}. Our goal is to
simulate losses $L$ for margins $F_1,\dots,F_d$ (assumed to be standard normal),
a given vector $\bm{w}=(w_1,\dots,w_d)$ of weights (assumed to be
$\bm{w}=(1,\dots,1)$), and different
\begin{itemize}
\item sample sizes $n$;
\item dimensions $d$;
\item copula families $C$ (note that we slightly abuse notation here and in what
  follows, using $C$ to denote a parametric copula family, not only a fixed copula); and
\item copula parameters, expressed in terms of the concordance measure
  Kendall's tau $\tau$,
\end{itemize}
and to compute $\VaR_\alpha(L)$ for different levels $\alpha$ (corresponding to the
Basel II/III rules for different risk types). This is a common setup and problem
from quantitative risk management. Since neither $F_L$, nor its quantile
function (and
thus $\VaR_\alpha(L)$) are known explicitly, we estimate $\VaR_\alpha(L)$ empirically
based on $n$ simulated losses $L_i$, $i\in\{1,\dots,n\}$, of $L$. This method
for estimating $\VaR_\alpha(L)$ is also known as \emph{Monte Carlo simulation
  method}; see \cite[Section~2.3.3]{mcneilfreyembrechts2005}. We repeat it $N_{sim}$ times
to be able to provide an error measure of the estimation via bootstrapped percentile confidence intervals.

\subsection[Translating the scientific problem to R]{Translating the scientific problem to \R}\label{sec:translate}
To summarize, our goal is to simulate, for each sample size $n$, dimension $d$,
copula family $C$, and strength of dependence
Kendall's tau $\tau$, $N_{sim}$ times $n$ losses $L_{ki}$, $k\in\{1,\dots,N_{sim}\}$,
$i\in\{1,\dots,n\}$, and to compute in the $k$th of the $N_{sim}$ replications
$\VaR_\alpha(L)$ as the empirical $\alpha$-quantile of $L_{ki}$,
$i\in\{1,\dots,n\}$, for each $\alpha$. Since different $\alpha$-quantiles can
(and should!) be estimated
based on the same simulated losses, we do not have to generate additional samples
for different values of $\alpha$, $\VaR_\alpha(L)$ can be estimated simultaneously
for all $\alpha$ under consideration.

% MM: require(simsalapar, lib="~/R/Pkgs/simsalapar_r1921-inst/"))

Table~\ref{tab:var} provides a summary of all variables involved in our
simulation study, their names in \R, \LaTeX\ expressions, type, and the
corresponding values we choose. Note that this table is produced entirely with
\pkg{simsalapar}'s \code{toLatex(varList, ....)}; see page~\pageref{lab:varList}.
\begin{table}[htbp]
  \centering
  \begin{tabular}{l*{2}{c}r}
    \toprule
    \multicolumn{1}{c}{Variable} & \multicolumn{1}{c}{expression} & \multicolumn{1}{c}{type} & \multicolumn{1}{c}{value} \\
    \midrule
    \texttt{n.sim} & \( N_{sim} \) & N & 32 \\
    \texttt{n} & \( n \) & grid & 64, 256 \\
    \texttt{d} & \( d \) & grid & 5, 20, 100, 500 \\
    \texttt{varWgts} & \( \mathbf{w} \) & frozen & 1, 1, 1, 1 \\
    \texttt{qF} & \( F ^ {- 1} \) & frozen & qF \\
    \texttt{family} & \( C \) & grid & Clayton, Gumbel \\
    \texttt{tau} & \( \tau \) & grid & 0.25, 0.50 \\
    \texttt{alpha} & \( \alpha \) & inner & 0.950, 0.990, 0.999 \\
    \bottomrule
  \end{tabular}
  \caption{Variables which determine our simulation study.}
  \label{tab:var}
\end{table}
For the moment, let us focus on the type. Available are:
\begin{itemize}[label=xxxxxxx]
\item[N:] The variable $N_{sim}$ gives the number of simulation (``bootstrap'')
  replications in our study.  This variable is present in many statistical
  simulations and allows one to provide an error measure of a statistical quantity
  such as an estimator. Because of this special meaning, it gets the type ``N'',
  and there can be only one variable of this type in a simulation study. If it is
  not given, it will implicitly be treated as 1.
\item[frozen:] The variable $\bm{w}$ is a list of length equal to the number of
  dimensions considered, where each entry is a vector (in our case a value which
  will be sufficiently often recycled by \R) of length equal to the
  corresponding dimension. Variables such as $\bm{w}$ (or the marginal quantile
  functions) remain the same throughout the whole simulation study, but one might
  want to change them if the study is conducted again. Variables of this type
  are assigned the type ``frozen'', since they remain fixed throughout the whole
  study.
\item[grid:] Variables of type ``grid'' are used to build a \emph{(physical)
    grid}. In \R\ this grid is implemented as a data frame. Each row in this
  data frame contains a unique combination of variables of type ``grid''. The
  number of rows $n_G$ of this grid, is thus the product of the lengths of all
  variables of type ``grid''. The simulation will iterate $N_{sim}$ times over all $n_G$
  rows and conduct the required computations. Conceptually, this corresponds to
  visiting each of the $N_{sim} \times n_G$ rows of a \emph{virtual grid} (seen
  as $N_{sim}$ copies of the grid pasted together). The computations for one row
  in this virtual grid are viewed as one \emph{sub-job}. In many situations,
  computing all sub-jobs sequentially turns out to be time-consuming (even after
  profiling of the code and removing time bombs such as deeply nested 'for'
  loops). In this situation, we can apply parallel computing and distribute the
  sub-jobs over several cores of a multi-core processor or several machines
  (nodes) in a cluster.
\item[inner:] Finally, variables of type ``inner'' are all dealt with within
  a sub-job for reasons of convenience, speed, load balancing etc. As mentioned
  before, in our example, $\alpha$ plays such a role since $\VaR_\alpha(L)$ can
  be estimated simultaneously for all $\alpha$ under consideration based on the
  same simulated losses.
\end{itemize}

As result of a simulation, we naturally obtain an array. This array has one
dimension for each variable of type ``grid'' or ``inner'', and one additional
dimension if $N_{sim}>1$. Besides the variable names, their type, and their
values, we also define \R\ expressions for each variable. These expressions are
later used to label tables or plots when the simulation results are analyzed.
\begin{remark}
  As an advantage of our approach based on \code{n.sim} in terms of
  load-balancing, each repeated simulation has the same expected run time. Note,
  however, that thousands of fast
  sub-jobs might lead to a comparably large overall run time due to both the
  waiting times for the jobs to start on a cluster and due to the overhead in
  communication between the master and the slaves. It might therefore be more
  efficient to send blocks of sub-jobs (say, 10 sub-jobs) to the same core or
  node. This feature is provided by the argument \code{block.size} in the
  \code{do*()} functions (\code{doLapply()}, \code{doForeach()},
  \code{doRmpi()}, \code{doMclapply()}, \code{doClusterApply()}) presented
  later.
\end{remark}

We are now ready to start writing an \R\ script which can be run on a
single computer or on a computer cluster. Since cluster types and interfaces are
quite different, we only focus on how to write the \R\ script here\footnote{As a
  quick example of how to run an \R\ script \file{simu.R} on different nodes on
  a computer cluster, let us briefly mention a specific example, the cluster
  \emph{Brutus} at ETH Zurich. It runs an LSF batch system. Once logged in, one
  can submit the script \file{simu.R} via \cmd{bsub -N -W 01:00 -n 48 -R
    "select[model==Opteron8380]" -R "span[ptile=16]" mpirun -n 1 R CMD BATCH
    simu.R}, for example, where the meaning of the various options is as
  follows: \cmd{-N} sends an email to the user when the batch job has finished;
  \cmd{-W 01:00} submits the job to the one-hour queue (jobs with this maximal
  wall-clock run time) on the cluster; the option \cmd{-n 48} asks for 48 cores
  (one is used as master, 47 as slaves); \cmd{-R
    "select[model==Opteron8380]"} % have 16 cores each
  specifies X86\_64 nodes with AMD Opteron 8380 CPUs for the sub-jobs to be run (this is
  important if run-time comparisons are required, since one has to make sure
  that the same architecture is used when computations are carried out in
  parallel); the option \cmd{-R "span[ptile=16]"} specifies that (all) 16 cores
  (on each node) are used on a single node (that means our job fully occupies $48/16=3$ nodes);
  \cmd{mpirun} specifies an Open MPI job which runs only one copy (\cmd{-n 1})
  of the program; and finally, \cmd{R CMD BATCH simu.R} is the standard call of
  the \R\ script \file{simu.R} in batch mode.}.
%% * After checking back with Olivier Byrde:
%% (see also http://brutuswiki.ethz.ch/brutus/Job_allocation_policy)
%% 1) A 'node' is a computer/machine in a cluster (possibly containing several
%%    processors/cores); above: node type is actually always "X86_64" but with
%%    our model-specification, we request Opteron8380-type processors (model)
%%    inside the node(s).
%% 2) Note: the only 48-core nodes are Opteron6174.
%% 3) Specifying ptile=3 would ask for 3 cores on each of the nodes
%%    => 16 nodes would start, each having 3 out of 16 cores working on our job
%%    Note: this is bad (always use 'all'), since the other 13 cores would
%%          be allocated to other jobs (affects the performance of your own job)
%% * Some more observations:
%%   + Running a slightly modified Gneiting demo on Brutus with the above call
%%     leads to:
%%     [hofertj@brutus2 ~]$ cat subjob-monitor.txt
%%     14:51:44 on a6346.hpc-net.ethz.ch: forecaster=statistician; value:  ...
%%     14:51:44 on a6292.hpc-net.ethz.ch: forecaster=   pessimist; value:  ...
%%     14:51:44 on a6294.hpc-net.ethz.ch: forecaster=    optimist; value:  ...
%%     => subjob-monitor.txt is correctly written in the directory from which
%%        the job is submitted
%%
%% * Note that i in  for(i in 1:N)  is *not* a function argument
%%   Principle:  some vars are *function* doCallWE()/doOne() arguments
%%   and some are not (N)
The first task is to implement the variable list presented above. Note that
\code{varlist()} is a generator for the S4 class \code{"varlist"}, which is only
little more than the usual \code{list()} in \R. For more details, use
\code{require(simsalapar)}, then \code{?varlist}, \code{getClass("varlist")}, or
\code{class?varlist}. Given a variable list of class \code{"varlist"}, a table
such as Table~\ref{tab:var} can be automatically generated with the
\code{toLatex.varlist} method.

\label{lab:varList}
\begin{Schunk}
\begin{Sinput}
> require("simsalapar")
> varList <- # *User provided* list of variables
      varlist( # constructor for an object of class 'varlist'
          ## replications
          n.sim = list(type="N", expr = quote(N[sim]), value = 32),
          ## sample size
          n = list(type="grid", value = c(64, 256)),
          ## dimensions, and weights (vector) for each d
          d = list(type="grid", value = c(5, 20, 100, 500)),
          varWgts = list(type="frozen", expr = quote(bold(w)),
                         value = list("5"=1, "20"=1, "100"=1, "500"=1)),
          ## margins
          qF = list(type="frozen", expr = quote(F^{-1}), value=list(qF=qnorm)),
          ## copula family names
          family=list(type="grid", expr = quote(C),
                      value = c("Clayton", "Gumbel")),
          ## dependencies by Kendall's tau
          tau = list(type="grid", value = c(0.25, 0.5)),
          ## levels corresponding to Basel II/III
          ## market risk (1d), market risk (10d), and credit risk, op.risk (1a)
          alpha = list(type="inner", value = c(0.95, 0.99, 0.999)))
> toLatex(varList, label = "tab:var",
          caption = "Variables which determine our simulation study.")
\end{Sinput}
\end{Schunk}

Note that one actually does not need to specify a type for \code{n.sim} or
variables of type ``frozen'', the default chosen is ``frozen'' unless the
variable is \code{n.sim} in which case it is ``N''.

The function \code{getEl()} can be used to extract elements of a certain type
from a variable list (defaults to all values).
\begin{Schunk}
\begin{Sinput}
> str(getEl(varList, "grid")) # extract "value" of variables of type "grid"
\end{Sinput}
\begin{Soutput}
List of 4
 $ n     : num [1:2] 64 256
 $ d     : num [1:4] 5 20 100 500
 $ family: chr [1:2] "Clayton" "Gumbel"
 $ tau   : num [1:2] 0.25 0.5
\end{Soutput}
\begin{Sinput}
> str(getEl(varList, "inner")) # extract "value" of variables of type "inner"
\end{Sinput}
\begin{Soutput}
List of 1
 $ alpha: num [1:3] 0.95 0.99 0.999
\end{Soutput}
\end{Schunk}

To have a look at the grid for our working example (containing all combinations of
variables of type ``grid''), the function \code{mkGrid()} can be used as follows.
\begin{Schunk}
\begin{Sinput}
> pGrid <- mkGrid(varList) # create *physical* (see below) grid
> str(pGrid)
\end{Sinput}
\begin{Soutput}
'data.frame':	32 obs. of  4 variables:
 $ n     : num  64 256 64 256 64 256 64 256 64 256 ...
 $ d     : num  5 5 20 20 100 100 500 500 5 5 ...
 $ family: chr  "Clayton" "Clayton" "Clayton" "Clayton" ...
 $ tau   : num  0.25 0.25 0.25 0.25 0.25 0.25 0.25 0.25 0.25 0.25 ...
\end{Soutput}
\end{Schunk}

\subsection{The result of a simulation}\label{sec:res}
Our route from here is to conduct the simulations required for each line of
the virtual grid (in parallel). As an important point, note that each computational \emph{result}
naturally consists of the following components:
\begin{itemize}[label={\code{.Random.seed}:}]
\item[\code{value}:] The actual value. This is can be a scalar, numeric vector, or numeric array whose dimensions depend on variables of type ``inner''. The computed entries also depend on variables of type ``frozen'', but they do not enter the result array as additional dimensions.
\item[\code{error}:] It is important to adequately track errors during
simulation studies. If one computation fails, we lose all results computed so
  far and thus have to do the work again (fix the error,
  move the files to the cluster, wait for the simulation job to start,
  wait for it to fail or to finish successfully in this next trial run etc.).
  To avoid this, we capture the errors to be able to deal with them after the simulation has been conducted.
  This also allows us to compute statistics about errors, such as percentages of runs producing errors etc.
\item[\code{warning}:] Similar to errors, warnings are important to catch. They
  may indicate non-convergence of an algorithm (or a maximal number of
  iterations reached etc.) and therefore impact reliability of the results.
\item[\code{time}:] Measured run time can also be an indicator of reliability in
  the sense that if computations are too fast/slow, there might be a programming
  error (not leading to an error or warning and thus being detected). For example, if one accidentally
  switches a logical condition, a large computation may return in almost no time
  because it simply ended up in the wrong case. If the value computed from this
  case is not suspicious, and if there were no warnings and errors, then
  run time is the only indicator of a possible bug in the code. Furthermore,
  measuring run time is also helpful for benchmarking and assessing the usefulness of a result (even if a computation or algorithm only runs
  sufficiently fast on a large cluster, it might not be suitable for a
  notebook and therefore might have limited use overall).
\item[\code{.Random.seed}:] The random seed right before the user-specified computations are carried out. This is useful for reproducing single results for
  debugging purposes.
\end{itemize}
In many simulation studies, also on an academic level, focus is put on \code{value} only. We therefore particularly stress all of these components, since they become more and more important for obtaining reliable
results the larger the conducted simulation study is. Furthermore, \code{error}, \code{warning}, and \code{.Random.seed} are important to consider especially during experimental stage of the simulation, for checking an implementation, and testing it for numerical stability.

The paradigm of \pkg{simsalapar} is that the user only has to take care of how to compute the \code{value} (the statistic the user is most interested in). All other components addressed above are automatically dealt with by \pkg{simsalapar}. We will come back to the latter in Section~\ref{sec:do}, after having thought about how to compute the \code{value} for our working example in the following section.

\subsection[Writing the problem-specific function doOne()]{Writing the problem-specific function \code{doOne()}}
Programming in \R\ is about writing \emph{functions}.
Our goal is now to write the workhorse of the simulation study: \code{doOne()}.
This function has to be designed for the particular simulation problem at hand and is
therefore given here (with Roxygen documentation) instead of being part of
\pkg{simsalapar}. \code{doOne()} computes the value (a numeric vector here) for the given arguments, that is, the component \code{value}. For functions \code{doOne()} for other simulations, we refer to the demos of \pkg{simsalapar}, see for example \code{demo(TGforecasts)} for reproducing the simulation conducted by \cite{gneiting2011}.
% Also, if required, \code{doOne()} could write an output file with intermediate results
% (for debugging, checking etc.).
% Robin: writing to a file is not so simple on nodes of a cluster...
%        possible solution: open a server, processes contact server;
%        at Google: global file system where each process has access
\begin{Schunk}
\begin{Sinput}
> ##' *User provided* function
> ##' @title Function to Compute the Results for One Line of the Virtual Grid
> ##' @param n sample size
> ##' @param d dimension
> ##' @param qF marginal quantile function
> ##' @param family copula family
> ##' @param tau Kendall's tau (determines strength of dependence)
> ##' @param alpha 'confidence' level alpha
> ##' @param varWgts vector of weights
> ##' @param names logical indicating whether the quantiles are named
> ##' @return value (vector of VaR_alpha(L) estimates for all alpha)
> ##' @author Marius Hofert and Martin Maechler
> doOne <- function(n, d, qF, family, tau, alpha, varWgts, names=FALSE)
  {
      ## checks (and load required packages here for parallel computing later on)
      w <- varWgts[[as.character(d)]]
      stopifnot(require(copula), # load 'copula'
                sapply(list(w, alpha, tau, d), is.numeric)) # sanity checks
  
      ## simulate risk-factor changes (if defined outside doOne(), use
      ## doOne <- local({...}) construction as in some of simsalapar's demos)
      simRFC <- function(n, d, qF, family, tau) {
          ## define the copula of the risk factor changes
          theta <- getAcop(family)@iTau(tau) # determine copula parameter
          cop <- onacopulaL(family, list(theta, 1:d)) # define the copula
          ## sample the meta-copula-model for the risk-factor changes X
          qF(rCopula(n, cop)) # simulate via Sklar's Theorem
      }
      X <- simRFC(n, d=d, qF=qF[["qF"]], family=family, tau=tau) # simulate X
  
      ## compute the losses and estimate VaR_alpha(L)
      L <- -rowSums(expm1(X) * matrix(rep(w, length.out=d),
                                      nrow=n, ncol=d, byrow=TRUE)) # losses
      quantile(L, probs=alpha, names=names) # empirical quantile as VaR estimate
  }
\end{Sinput}
\end{Schunk}

% basic check (hidden)

\subsection[Putting the pieces together: The do*() functions]{Putting the pieces together: The \code{do*()} functions}\label{sec:do}
\begin{figure}[htbp]% onion
  \centering\small
  \begin{tikzpicture}[decoration=zigzag, scale=0.9]% idea: major axis +1cm, minor axis +0.9cm
    %% largest ellipse
    \filldraw[fill=gray!10, draw=gray!80] (0,0) ellipse (5cm and 2.8cm);
    \path[postaction={decorate,
      decoration={
        raise=-1.4em,
        text along path,
        text={|\tt|doLapply(), ..., doMclapply(), doClusterApply()},
        text align=center,
      },
    }] (0,0) ++(160:5cm and 2.8cm) arc (160:20:5cm and 2.8cm);
    %% 2nd largest ellipse
    \filldraw[fill=gray!20, draw=gray!80] (0,0) ellipse (4cm and 2cm);
    \path[postaction={decorate,
      decoration={
        raise=-1.4em,
        text along path,
        text={|\tt|subjob()},
        text align=center,
      },
    }] (0,0) ++(160:4cm and 2cm) arc (160:20:4cm and 2cm);
    %% 3rd largest ellipse
    \filldraw[fill=gray!30, draw=gray!80] (0,0) ellipse (3cm and 1.2cm);
    \path[postaction={decorate,
      decoration={
        raise=-1.4em,
        text along path,
        text={|\tt|doCallWE()},
        text align=center,
      },
    }] (0,0) ++(160:3cm and 1.2cm) arc(160:20:3cm and 1.2cm);
    %% smallest ellipse
    \filldraw[fill=blue!40, draw=gray!80] (0,0) ellipse (2cm and 0.4cm) node
    {\texttt{doOne()}};
  \end{tikzpicture}
  \caption{Layers of functions involved in a simulation
    study. \pkg{simsalapar} provides all but \code{doOne()}.}
  \label{fig:functions}
\end{figure}
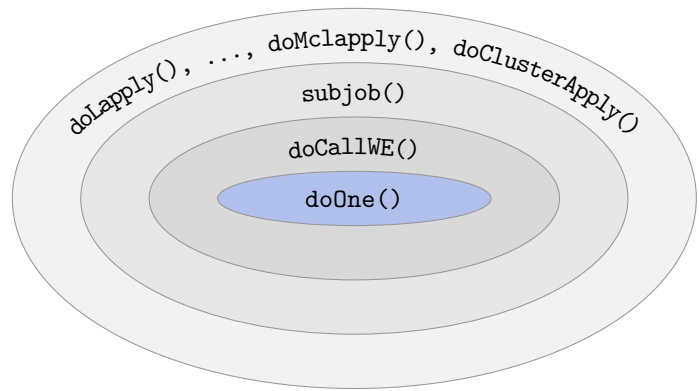

To conduct the main simulation, we only need one more function which
iterates over all sub-jobs and calls \code{doOne()}. There are several options:
sequential (see Section~\ref{sec:lapply}) versus various approaches for parallel
computing (see Section~\ref{sec:parallel}), for which we provide the
\code{do*()} functions explained below. Since these functions are quite
technical and lengthy, we will present the details in Section~\ref{sec:behind}.
For the moment, our goal is to understand the functions they call
in order to understand how the simulation works. Figure~\ref{fig:functions} visualizes the main functions involved in conducting the simulation.
% Note:
% - doCallWE() also calls tryCatch.W.E() and mkTimer()
% - do*() functions call mkGrid(), get.nonGrids(), saveSim();
%   saveSim() essentially calls mkAL()
These functions break down the whole task into smaller pieces (which
improves readability of the code and simplifies debugging when procedures
fail).

We have already discussed the innermost, user-provided function
\code{doOne()}. The auxiliary function \code{doCallWE()} captures the values
computed by \code{doOne()} (or \code{NULL} if there was an error), errors (or
\code{NULL} if there was no error), warnings (or \code{NULL} if there was no
warning), and run times when calling \code{doOne()} (by default user time in
milliseconds without garbage collection in order to save time; see
\code{mkTimer()}; for serious run time measurement, use \code{timer =
  mkTimer(gcFirst=TRUE)} in \code{doCallWE()}). % and a *robust* analysis (to remove the influence of require() etc.)
For details about how \code{doCallWE()} achieves this (and thus an explanation
for its name), see Section~\ref{sec:select:sim}. This already provides us with a
list of four of the five components of a result as addressed in
Section~\ref{sec:res}. The component \code{.Random.seed} may\footnote{\code{subjob}'s
  default \code{keepSeed=FALSE} has been chosen to avoid large result objects.}
then be added by the function which calls \code{doCallWE()}, namely
\code{subjob()}. The aim of \code{subjob()} is to compute one sub-job, that is,
one row of the virtual grid. A large part of this function deals with correctly
setting the seed. It also provides a monitor feature; see
Section~\ref{sec:select:sim} for the details.

As mentioned before, there are several choices available for the outermost layer
of functions, depending on whether, and if yes, what kind of
parallel computing should be used to deal with the rows of the virtual grid. In
particular, \pkg{simsalapar} provides the following functions, see Section~\ref{sec:behind}:
\begin{itemize}[label={\code{doClusterApply()}:}]\label{doFOOapply-list}
\item[\code{doLapply()}:] a wrapper for the non-parallel function
  \code{lapply()}. This is useful for testing the code with a small number of
  different parameters so that the simulation still runs locally on the
  computer at hand.
\item[\code{doForeach()}:] a wrapper for the function \code{foreach()} of the
  \R\ package \pkg{foreach} to conduct computations in parallel on several cores
  or nodes. A version specific to our working example based on nested
  \code{foreach()} loops is presented in Section~\ref{sec:behind}.
\item[\code{doRmpi()}:] a wrapper for the function \code{mpi.apply()} or its
  load-balancing version \code{mpi.applyLB()} (default) from the \R\ package
  \pkg{Rmpi} for parallel computing on several cores or nodes.
\item[\code{doMclapply()}:] a wrapper for the function \code{mclapply()} (with
  (default) or without load-balancing) of the \R\ package
  \pkg{parallel} for parallel computing on several cores (not working on Windows).
\item[\code{doClusterApply()}:] a wrapper for the function \code{clusterApply()} or its
  load-balancing version \code{clusterApplyLB()} (default) of the \R\ package \pkg{parallel}
  for parallel computing on several cores or nodes.
\end{itemize}

\begin{remark}
  The user of \pkg{simsalapar} can call one of the above functions \code{do*()}
  to finally run the whole simulation study; see Sections \ref{sec:lapply} and
  \ref{sec:parallel}. To this end, these functions iterate over all sub-jobs and
  finally call the function \code{saveSim()}; see
  Section~\ref{sec:select:sim}. \code{saveSim()} tries to convert the resulting
  list of lists of length four or five to an array of lists of length four or
  five and saves it in the \cmd{.rds} file specified by the argument
  \code{sfile}. If this non-trivial conversion fails\footnote{Our flexible
    approach allows one to implement a function \code{doOne()} such that the order in
    which the ``inner'' variables appear does not correspond to the order in
    which they appear in the variable list. Therefore, the user-provided
    workhorse \code{doOne()} has to be written with care.},
  %% Although the functions \code{do*()} % actually \code{mkAL()} behind \code{saveSim()}
  %% conducts basic checks, there is no guarantee that the resulting array is
  %% correctly filled (and most likely it will not be) unless the user provides a
  %% return value object which respects the order of the ``inner'' variables as
  %% specified in \code{varList}.
  the raw list of lists of length four or five is saved instead,
  so that results are not lost. This behavior can also be obtained by directly
  specifying \code{doAL=FALSE} when calling the \code{do*()} functions. To
  further avoid that the conversion fails, the functions \code{do*()} conduct a
  basic check of the correctness of the return value of \code{doOne()} by
  calling the function \code{doCheck()}. This can also be called by the user
  after implementing \code{doOne()} to verify the correctness of \code{doOne()}; see, for example,
  \code{demo(VaRsuperadd)}.
\end{remark}

\subsection[Running the simulation sequentially: doLapply() based on lapply()]{%
  Running the simulation sequentially: \code{doLapply()} based on
  \code{lapply()}}\label{sec:lapply}
In Sections~\ref{sec:parallel} and \ref{sec:behind}, we will compare different
approaches for parallel computing in \R. To make this easier to follow, we start
with \code{doLapply()}, see Section~\ref{sec:select:sim}, which is a wrapper for
the sequential (non-parallel) function \code{lapply()} to iterate over all
rows of the virtual grid. This sequential approach is often the first choice to
try (for a smaller number of parameter combinations) in order to check whether the
simulation actually does what it should, for debugging etc. If sequential
computations based on \code{lapply()} turn out to be too slow, one can easily use one
of the parallel computing approaches described in Sections~\ref{sec:parallel}
and \ref{sec:behind}, since they share the same interface.

We now demonstrate the use of \code{doLapply()} to run the whole
simulation. Note that \code{names} is an optional argument to our \code{doOne()}
and the argument \code{monitor}, passed to \code{subjob()}, allows progress monitoring.
\begin{Schunk}
\begin{Sinput}
> ## our working example
> res <- doLapply(varList, sfile="res_lapply_seq.rds", doOne=doOne, names=TRUE,
                  monitor=interactive())
\end{Sinput}
\end{Schunk}

The \code{str()}ucture of the resulting object can be briefly analyzed as
follows (note that the dimension for \code{n.sim} is not named, thus \code{dimnames(res)$n.sim} is \code{NULL}).
\begin{Schunk}
\begin{Sinput}
> str(res, max.level=2)
\end{Sinput}
\end{Schunk}
\begin{Schunk}
\begin{Soutput}
List of 1024
 $ :List of 4
  ..$ value  : num [1:3(1d)] 3.18 3.6 4.02
  .. ..- attr(*, "dimnames")=List of 1
  ..$ error  : NULL
  ..$ warning: NULL
  ..$ time   : num 21
 $ :List of 4
  ..$ value  : num [1:3(1d)] 3.36 4.35 4.68
  .. ..- attr(*, "dimnames")=List of 1
  ..$ error  : NULL
  ..$ warning: NULL
  ..$ time   : num 1
 ....... 
 ....... 
  [list output truncated]
 - attr(*, "dim")= Named int [1:5] 2 4 2 2 32
  ..- attr(*, "names")= chr [1:5] "n" "d" "family" "tau" ...
 - attr(*, "dimnames")=List of 5
  ..$ n     : chr [1:2] "64" "256"
  ..$ d     : chr [1:4] "5" "20" "100" "500"
  ..$ family: chr [1:2] "Clayton" "Gumbel"
  ..$ tau   : chr [1:2] "0.25" "0.50"
  ..$ n.sim : NULL
 - attr(*, "fromFile")= logi TRUE
\end{Soutput}
\end{Schunk}
\begin{Schunk}
\begin{Sinput}
> str(dimnames(res))
\end{Sinput}
\begin{Soutput}
List of 5
 $ n     : chr [1:2] "64" "256"
 $ d     : chr [1:4] "5" "20" "100" "500"
 $ family: chr [1:2] "Clayton" "Gumbel"
 $ tau   : chr [1:2] "0.25" "0.50"
 $ n.sim : NULL
\end{Soutput}
\end{Schunk}
% We will call \code{doLapply()} in a few other setups in Section~\ref{sec:behind:varlists}, including different seeding methods, etc.

%%% Local Variables:
%%% TeX-master: "parallel.tex"
%%% End:

% patchDVI setup
\Sconcordance{concordance:03_parallel_R.tex:03_parallel_R.Rnw:%
1 6 1 1 5 21 1 1 3 5 0 1 8 3 1 1 7 4 0}

% Sweave options
 % use pdfcrop to crop .pdf
 % put figures in ./ and name them with prefix "fig-"
%\VignetteDepends{simsalapar}
%\VignetteDepends{copula}

\section[Parallel computing in R]{Parallel computing in \R}\label{sec:parallel}
In the same way that \code{doLapply()} wraps around \code{lapply()},
\pkg{simsalapar} provides convenient wrapper functions
to conduct the \emph{same} computations (but) \emph{in parallel}.
These different approaches are useful for different kinds of setups,
such as different available computer architectures or different specifications of the
simulation study considered. Before we go into the details, let us mention that one
should only use \emph{one} of the \code{do*()} functions. Mixing several different ways of
conducting parallel computations in the same \R\ process might lead to weird
errors, conflicts of various kinds, or unreliable results at best.
% for example "Error: 'recvOneData.MPIclusted' is not an exported object from 'namespace:snow'" when running foreach after the other cases

For conducting computations in parallel with \R, one just needs to replace
\code{doLapply()} above (Section~\ref{sec:lapply}) by one of its
\emph{``parallelized''} \code{do*()} versions listed in
Section~\ref{sec:do}. We will
take \code{doClusterApply()} as an example here and refer to
Section~\ref{sec:behind} for a more in-depth analysis and comparison of the
results obtained from these different approaches to those from
\code{doLapply()} to check their correctness, consistency, and efficiency. % this is a promise !

\begin{Schunk}
\begin{Sinput}
> res5 <- doClusterApply(varList, sfile="res5_clApply_seq.rds",
                         doOne=doOne, names=TRUE)
\end{Sinput}
\end{Schunk}
% Redo only this one: \rm res5_clApply_seq.rds 03_parallel_R.tex 06_wrapup.tex ; make
%
Indeed, \code{doClusterApply()} produces the same result as
\code{doLapply()} did above:
\begin{Schunk}
\begin{Sinput}
> stopifnot(doRes.equal(res5, res)) # note: doRes.equal() is part of simsalapar
\end{Sinput}
\end{Schunk}

% patchDVI setup
\Sconcordance{concordance:04_analysis.tex:04_analysis.Rnw:%
1 6 1 1 6 32 1 1 2 1 0 3 1 3 0 1 2 1 1 1 2 1 0 1 1 13 0 1 4 4 %
1 1 2 1 0 2 1 15 0 1 1 16 0 1 2 2 1 1 2 16 0 1 2 1 1 20 0 1 2 %
24 1 1 2 1 0 3 1 3 0 1 2 6 1 1 3 2 0 1 1 1 2 1 0 2 1 1 2 1 0 1 %
2 1 0 1 1 3 0 1 2 4 1 1 2 4 0 1 2 1 1 1 5 7 0 1 3 27 0 1 2 84 %
1 1 2 1 0 1 2 1 0 1 2 5 0 1 2 5 1 1 2 1 0 1 1 1 3 6 0 1 2 5 1 %
1 65 3 1}

% Sweave options
 % use pdfcrop to crop .pdf
 % put figures in ./ and name them with prefix "fig-"
%\VignetteDepends{simsalapar}
%\VignetteDepends{copula}

\section{Data Analysis}\label{sec:analysis}
After having conducted the main simulation, the final task is to analyze the
data and present the results. It seems difficult to provide a general solution for this
part of the simulation study. Besides the solutions provided by \pkg{simsalapar}
however, it might therefore be required to write additional
problem-specific functions. In this case, functions from \pkg{simsalapar}
may at least serve as good starting points.

The function \code{getArray()}, presented in Section~\ref{sec:select:ana}, is a
function from \pkg{simsalapar} which, given the result object of the simulation
and one of the components ``value'' (the default), ``error'', ``warning'', or
``time'' creates an array containing the corresponding results. This is
typically more convenient than working with an array of lists, which the object
as returned by one of the \code{do*()} functions naturally is. For the
components being ``error'' or ``warning'', the array created contains (by
default) boolean variables indicating whether there was an error or warning,
respectively. This behavior can be changed by providing a suitable argument
\code{FUN} to \code{getArray()}. Additionally, \code{getArray()} allows for an
argument \code{err.value}, defaulting to \code{NA}, for replacing values in case
there was an error. As mentioned before, each ``value'', can be a scalar, a
numeric vector, or a numeric array, often with \code{dimnames}, e.g., resulting
from (the outer product of) variables of type ``inner''. Note that for
conducting the simulation, variables sometimes can be declared as ``inner'' or
``frozen'' interchangeably. However, this changes the dimension of the result
object for the analysis in the sense that variables of type ``inner'' appear as
additional dimensions in the result array and can thus serve as a proper quantity/dimension in a
table or plot, whereas variables of type ``frozen'' do not.

Since it is the most compatible across different architectures (if the reader
wants to reproduce our results), we consider the result object \code{res} as returned by
\code{doLapply()} here. For our working example, we can apply \code{getArray()}
to \code{res} as follows.
\begin{Schunk}
\begin{Sinput}
> val  <- getArray(res) # array of values
> err  <- getArray(res, "error") # array of error indicators
> warn <- getArray(res, "warning") # array of warning indicators
> time <- getArray(res, "time") # array of user times in ms
\end{Sinput}
\end{Schunk}
If we wanted, we now could base all further analysis on a \code{data.frame}
which is easily produced from our array of values via \code{array2df()}:
\begin{Schunk}
\begin{Sinput}
> df <- array2df(val)
> str(df)
\end{Sinput}
\begin{Soutput}
'data.frame':	3072 obs. of  7 variables:
 $ alpha : Factor w/ 3 levels "95%","99%","99.9%": 1 2 3 1 2 3 1 2 3 1 ...
 $ n     : Factor w/ 2 levels "64","256": 1 1 1 2 2 2 1 1 1 2 ...
 $ d     : Factor w/ 4 levels "5","20","100",..: 1 1 1 1 1 1 2 2 2 2 ...
 $ family: Factor w/ 2 levels "Clayton","Gumbel": 1 1 1 1 1 1 1 1 1 1 ...
 $ tau   : Factor w/ 2 levels "0.25","0.50": 1 1 1 1 1 1 1 1 1 1 ...
 $ n.sim : Factor w/ 32 levels "1","2","3","4",..: 1 1 1 1 1 1 1 1 1 1 ...
 $ value : num  3.18 3.6 4.02 3.36 4.35 ...
\end{Soutput}
\end{Schunk}

As a first part of the analysis, we are interested in how reliable our results
are. We thus consider possible errors and warnings of the computations
conducted. Flat contingency tables (obtained by \code{ftable()}) allow us to
conveniently get an overview as follows.
\begin{Schunk}
\begin{Sinput}
> rv <- c("family", "d") # row variables
> cv <- c("tau", "n") # column variables
> ftable(100* err, row.vars = rv, col.vars = cv) # % of errors
\end{Sinput}
\begin{Soutput}
            tau 0.25     0.50    
            n     64 256   64 256
family  d                        
Clayton 5          0   0    0   0
        20         0   0    0   0
        100        0   0    0   0
        500        0   0    0   0
Gumbel  5          0   0    0   0
        20         0   0    0   0
        100        0   0    0   0
        500        0   0    0   0
\end{Soutput}
\begin{Sinput}
> ftable(100*warn, row.vars = rv, col.vars = cv) # % of warnings
\end{Sinput}
\begin{Soutput}
            tau 0.25     0.50    
            n     64 256   64 256
family  d                        
Clayton 5          0   0    0   0
        20         0   0    0   0
        100        0   0    0   0
        500        0   0    0   0
Gumbel  5          0   0    0   0
        20         0   0    0   0
        100        0   0    0   0
        500        0   0    0   0
\end{Soutput}
\end{Schunk}

Since we neither have warnings nor errors in our numerically non-critical
example study, let us briefly consider the run times:
\begin{Schunk}
\begin{Sinput}
> ftable(time, row.vars = rv, col.vars = cv) # run times
\end{Sinput}
\begin{Soutput}
            tau 0.25      0.50     
            n     64  256   64  256
family  d                          
Clayton 5         86   91   66   85
        20        87  157   92  155
        100      180  517  175  522
        500      636 3259  621 3190
Gumbel  5         73   98   72   94
        20        93  176   96  171
        100      193  584  192  577
        500      922 3244  860 3344
\end{Soutput}
\begin{Sinput}
> dtime <- array2df(time)
> summary(dtime)
\end{Sinput}
\begin{Soutput}
   n         d           family      tau          n.sim    
 64 :512   5  :256   Clayton:512   0.25:512   1      : 32  
 256:512   20 :256   Gumbel :512   0.50:512   2      : 32  
           100:256                            3      : 32  
           500:256                            4      : 32  
                                              5      : 32  
                                              6      : 32  
                                              (Other):832  
     value       
 Min.   :  0.00  
 1st Qu.:  3.00  
 Median :  5.00  
 Mean   : 20.22  
 3rd Qu.: 19.00  
 Max.   :302.00  
\end{Soutput}
\end{Schunk}

In what follows, we exclusively focus on the actual computed values, hence
the array \code{val}. We apply tools from \pkg{simsalapar} that allow us to
create flexible \LaTeX\ tables and sophisticated graphs for representing these
results.

\subsection[Creating LaTeX tables]{Creating \LaTeX\ tables}
In this section, we create \LaTeX\ tables of the results. Our goal is to make
this process modular and flexible. We thus leave tasks such as
formatting of table entries as much as possible to the user. Note that there are
already \R\ packages available for generating \LaTeX\ tables, for example the
well-known \pkg{xtable} or the rather new \pkg{tables}. However, they do not
fulfill the above requirements (and come with other unwanted side effects
concerning the table headers or formatting of entries we do not want to cope
with). We therefore present new tools for constructing tables with
\pkg{simsalapar}. For inclusion in \LaTeX\ documents, only the \LaTeX\ package
\pkg{tabularx}, and, due to our defaults following the paradigm of
\pkg{booktabs}, the \LaTeX\ package \pkg{booktabs} have to be loaded in the \cmd{.tex}
document. Much more sophisticated alignment of column entries for \LaTeX\ tables
than we show here (even including units) can be achieved in combination with the
\LaTeX\ package \pkg{siunitx}; see its corresponding extensive manual. Note that
these packages all come with standard \LaTeX\ distributions.

After having computed arrays of (robust) Value-at-Risk estimates and (robust)
standard deviations via
\begin{Schunk}
\begin{Sinput}
> non.sim.margins <- setdiff(names(dimnames(val)), "n.sim")
> huber. <- function(x) MASS::huber(x)$mu # or better  robustbase::huberM(x)$mu
> VaR     <- apply(val, non.sim.margins, huber.) # (robust) VaR estimates
> VaR.mad <- apply(val, non.sim.margins, mad) # median absolute deviation
\end{Sinput}
\end{Schunk}
we format and merge the arrays. As just mentioned, we specifically leave this
task to the user to guarantee flexibility. As an example, we put the
(robust) standard deviations in parentheses and colorize\footnote{This
  requires the \LaTeX\ package \pkg{xcolor} with the option \code{table} to be
  loaded in the \LaTeX\ document. The latter option even allows to use
  \code{$\backslash$cellcolor} to modify the background colors of select table
  cells.} all entries corresponding to the largest level $\alpha$.
\begin{Schunk}
\begin{Sinput}
> ## format values and mads
> fval <- formatC(VaR, digits=1, format="f")
> fmad <- paste0("(", format(round(VaR.mad, 1), scientific=FALSE, trim=TRUE), ")")
> ## paste together
> nc <- nchar(fmad)
> sm <- nc == min(nc) # indices of smaller numbers
> fmad[sm] <- paste0("\\ \\,", fmad[sm])
> fres <- array(paste(fval, fmad), # paste the results together
                dim=dim(fval), dimnames=dimnames(fval))
> ## colorize entries
> ia <- dim(fval)[1] # index of largest alpha
> fres[ia,,,,] <- paste("\\color{white!40!black}", fres[ia,,,,])
\end{Sinput}
\end{Schunk}
%% for removing NA, NaN, use: fval[is.na(fval)] <- ""

Next, we create a flat contingency table from the array of formatted results \code{fres}. The arguments
\code{row.vars} and \code{col.vars} of \code{ftable()} specify the basic layout
of Table~\ref{tab:ft} below.
\begin{Schunk}
\begin{Sinput}
> ft <- ftable(fres, row.vars=c("family","n","d"), col.vars=c("tau","alpha"))
\end{Sinput}
\end{Schunk}

Table~\ref{tab:ft} shows the results.
\begin{Schunk}
\begin{Sinput}
> tabL <- toLatex(ft, vList = varList,
                  fontsize = "scriptsize",
                  caption = "Table of results constructed with the \\code{ftable} method \\code{toLatex.ftable}.",
                  label = "tab:ft")
\end{Sinput}
\end{Schunk}
\begin{table}[htbp]
  \centering\scriptsize
  \begin{tabular}{*{3}{l}*{6}{r}}
    \toprule
     &  & \( \tau \) & \multicolumn{3}{c}{0.25} & \multicolumn{3}{c}{0.50} \\
    \cmidrule(lr){4-6} \cmidrule(lr){7-9}
    \( C \) & \( n \) & \( d \) \textbar\ \( \alpha \) & \multicolumn{1}{c}{95\%} & \multicolumn{1}{c}{99\%} & \multicolumn{1}{c}{99.9\%} & \multicolumn{1}{c}{95\%} & \multicolumn{1}{c}{99\%} & \multicolumn{1}{c}{99.9\%} \\
    \midrule
    Clayton & 64 & 5 & 3.1 \ \,(0.4) & 3.8 \ \,(0.4) & \color{white!40!black} 4.0 \ \,(0.5) & 3.6 \ \,(0.3) & 4.2 \ \,(0.2) & \color{white!40!black} 4.4 \ \,(0.2) \\
    &  & 20 & 10.6 \ \,(1.4) & 13.5 \ \,(1.5) & \color{white!40!black} 14.8 \ \,(2.2) & 14.2 \ \,(1.6) & 16.7 \ \,(1.0) & \color{white!40!black} 17.4 \ \,(1.0) \\
    &  & 100 & 46.1 \ \,(9.1) & 63.5 (11.6) & \color{white!40!black} 68.5 (13.6) & 70.7 \ \,(8.6) & 83.7 \ \,(3.9) & \color{white!40!black} 86.7 \ \,(4.2) \\
    &  & 500 & 224.8 (50.6) & 307.8 (61.5) & \color{white!40!black} 336.0 (66.8) & 350.0 (40.5) & 418.6 (22.3) & \color{white!40!black} 434.0 (21.4) \\ \addlinespace[3pt]
    & 256 & 5 & 3.2 \ \,(0.2) & 4.1 \ \,(0.2) & \color{white!40!black} 4.4 \ \,(0.2) & 3.9 \ \,(0.2) & 4.4 \ \,(0.1) & \color{white!40!black} 4.6 \ \,(0.1) \\
    &  & 20 & 10.9 \ \,(1.0) & 15.3 \ \,(1.2) & \color{white!40!black} 17.0 \ \,(0.9) & 15.3 \ \,(0.7) & 17.6 \ \,(0.5) & \color{white!40!black} 18.5 \ \,(0.6) \\
    &  & 100 & 49.0 \ \,(5.5) & 72.1 \ \,(7.7) & \color{white!40!black} 82.5 \ \,(4.8) & 76.0 \ \,(3.4) & 87.9 \ \,(2.7) & \color{white!40!black} 92.3 \ \,(3.0) \\
    &  & 500 & 240.4 (27.0) & 349.7 (35.3) & \color{white!40!black} 408.5 (24.3) & 378.8 (17.4) & 439.4 (12.7) & \color{white!40!black} 461.7 (14.2) \\ \addlinespace[6pt]
    Gumbel & 64 & 5 & 2.7 \ \,(0.3) & 3.3 \ \,(0.4) & \color{white!40!black} 3.4 \ \,(0.5) & 3.3 \ \,(0.3) & 3.8 \ \,(0.3) & \color{white!40!black} 4.0 \ \,(0.2) \\
    &  & 20 & 7.3 \ \,(1.1) & 9.4 \ \,(1.2) & \color{white!40!black} 10.1 \ \,(1.5) & 12.2 \ \,(0.6) & 14.0 \ \,(1.2) & \color{white!40!black} 14.6 \ \,(1.2) \\
    &  & 100 & 26.0 \ \,(4.2) & 35.8 \ \,(4.7) & \color{white!40!black} 38.5 \ \,(5.6) & 57.7 \ \,(5.1) & 67.7 \ \,(4.8) & \color{white!40!black} 70.3 \ \,(5.4) \\
    &  & 500 & 117.2 (12.5) & 154.4 (19.0) & \color{white!40!black} 167.5 (18.2) & 288.2 (18.0) & 333.7 (23.0) & \color{white!40!black} 347.9 (20.7) \\ \addlinespace[3pt]
    & 256 & 5 & 2.7 \ \,(0.2) & 3.3 \ \,(0.2) & \color{white!40!black} 3.7 \ \,(0.2) & 3.4 \ \,(0.2) & 3.9 \ \,(0.1) & \color{white!40!black} 4.2 \ \,(0.1) \\
    &  & 20 & 7.4 \ \,(0.5) & 9.9 \ \,(0.8) & \color{white!40!black} 11.5 \ \,(0.9) & 12.5 \ \,(0.4) & 14.7 \ \,(0.7) & \color{white!40!black} 16.0 \ \,(0.6) \\
    &  & 100 & 27.8 \ \,(2.8) & 38.4 \ \,(3.1) & \color{white!40!black} 44.7 \ \,(3.2) & 60.4 \ \,(2.3) & 70.9 \ \,(2.5) & \color{white!40!black} 76.9 \ \,(3.5) \\
    &  & 500 & 126.8 (10.3) & 171.9 (11.2) & \color{white!40!black} 202.3 (13.5) & 299.1 (13.7) & 353.8 (13.2) & \color{white!40!black} 380.0 \ \,(9.7) \\
    \bottomrule
  \end{tabular}
  \caption{Table of results constructed with the \code{ftable} method \code{toLatex.ftable}.}
  \label{tab:ft}
\end{table}
To summarize, using functions from \pkg{simsalapar} and packages from \LaTeX,
one can create flexible \LaTeX\ tables. If the simulation results become
sufficiently complicated, creating \LaTeX\ tables (or at least parts of them)
from \R\ reduces a lot of work, especially if the simulation study has to be
repeated due to bug fixes, improvements, or changes in the implementation. Note
that the table header typically constitutes the main complication when
constructing tables. It might still require manual modifications in case our
carefully chosen defaults do not suffice. \pkg{simsalapar} provides many other
functions not presented here, including the (currently non-exported) functions
\code{ftable2latex()} and \code{fftable()} and the (exported) functions
\code{tablines()} and \code{wrapLaTable()}. These ingredient functions of the
method \code{toLatex.ftable} can still be useful if one encounters very specific
requirements not covered by \code{toLatex.ftable}. More details on the latter
can be found in Section~\ref{sec:select:ana}. A crucial step in the development
of \code{tablines()} was the correct formatting of an \code{ftable} without
introducing empty rows or columns. For this we introduced four different methods
of ``compactness'' of a formatted \code{ftable} which are available in
\code{format.ftable()} from \R\ version 3.0.0 and for earlier versions in
\pkg{simsalapar}.

\subsection{Graphical analysis}
Next we show how \pkg{simsalapar} can be applied to visualize the results of our
study. In modern statistics, displaying results with graphics (as opposed to
tables) is typically good practice, since it is easier to see the story the data
would like to tell us. For example, in a table, the human eye can only compare
two numbers at a time, in well-designed graphics much more information
is visible.

There are various different approaches of how to create graphics in \R, for
example, with the traditional \pkg{graphics} package, the \pkg{lattice}, or the
\pkg{ggplot2} package. The most flexible approach is based on \pkg{grid}
graphics; see \cite{murrell2006}. In what follows, we apply the function
\code{mayplot()} (based on \pkg{grid} and \pkg{graphics} via \pkg{gridBase})
from \pkg{simsalapar} for creating a plot matrix (also known as
\emph{conditioning plot}) from an array of values. Within each cell of this plot
a traditional graphic is drawn to visualize the
results.
%% Note that matrix-like plots of this type can also be constructed with the
%% packages \pkg{lattice} or \pkg{ggplot2}. Since both packages come at the price
%% of limitations we do not like to cope with, we use the more flexible approach
%% based on \pkg{grid} and \pkg{gridBase}.

In our example study, the strength of dependence in terms of
Kendall's tau determines the columns of the matrix-like plot and the copula
family determines its rows. In each cell, there is an x and a
y axis. For making comparisons easier, one typically would like to have the same
limits on the y axes across different rows of the plot matrix. Sometimes it
makes sense to have separate scales for y axes in different rows (while still having the same
scales for all plots within the same row). This behavior can be determined with the
argument \code{ylim} (being \code{"global"} (the default) or \code{"local"}) of
\code{mayplot()}. For our working example, the x axis provides the different significance
levels $\alpha$. We thus naturally can depict three different input variables in
such a layout (copula families, Kendall's taus, and significance levels
$\alpha$). The y axis may show point estimates or boxplots of the simulated
Value-at-Risk values as given in \code{val}.

All other variables (sample sizes $n$, dimensions $d$) then have to be depicted
in the same cell, visually distinguished by different line types or colors, for
example (currently one such variable is allowed; we chose $d$ below by fixing
$n=256$). If more variables are involved, one might even want to put more
variables in one cell, rethink the design, or split different values of a
variable over separate plots. $N_{sim}$, if available, enters the scene through
a second label on the right side of graphic.

\begin{figure}[htb!]
% (width, height): large enough to make fonts small
\begin{Schunk}
\begin{Sinput}
> v256 <- val[, n = "256",,,,] # data to plot; alpha, d, family, tau, 1:n.sim
> ## adjust tau labels:
> dimnames(v256)[["tau"]] <- paste0("tau==", dimnames(v256)[["tau"]])
> mayplot(v256, varList, row.vars="family", col.vars="tau", xvar="alpha",
          ylab = bquote(widehat(VaR)[alpha](italic(L)))) # uses default xlab
\end{Sinput}
\end{Schunk}
\includegraphics{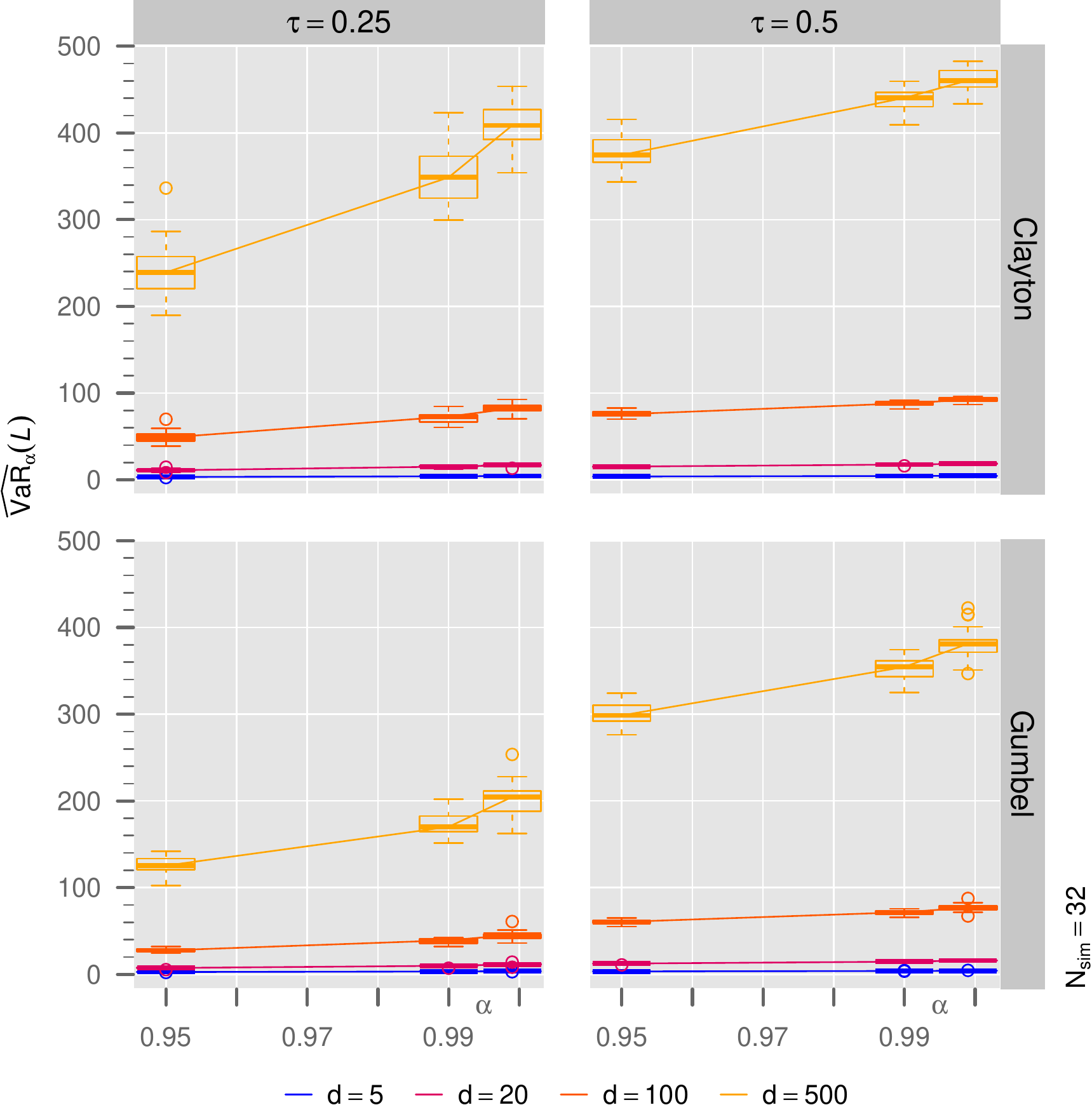}
\caption{Boxplots of the $N_{sim}$ simulated $\VaR_\alpha(L)$ values for $n=256$.}
\label{fig:VaR-256-box}
\end{figure}

\begin{figure}[htb!]
\begin{Schunk}
\begin{Sinput}
> varList. <- set.n.sim(varList, 1) # set n.sim=1 to get (default) lines plot
> dimnames(VaR)[["tau"]] <- paste0("tau==", dimnames(VaR)[["tau"]])
> mayplot(VaR[,n="256",,,], varList., row.vars="family", col.vars="tau",
          xvar="alpha", type = "b", log = "y", axlabspc = c(0.15, 0.08),
          ylab = bquote(widehat(VaR)[alpha](italic(L))))
\end{Sinput}
\end{Schunk}
\includegraphics{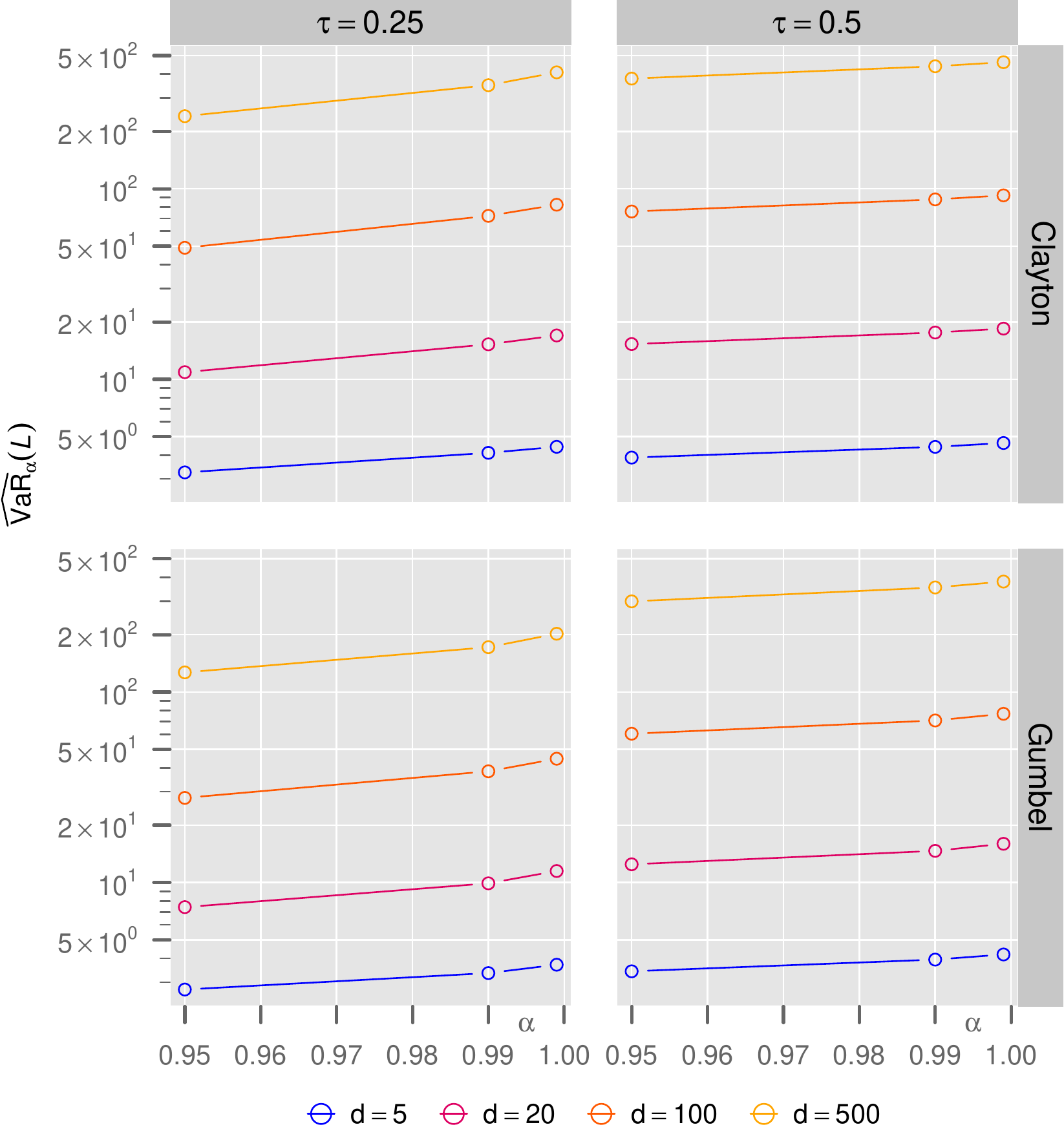}
\caption{Plot of robust $\VaR_\alpha(L)$ estimates in log scale, i.e., Huber
  ``means'' of $N_{sim}$ values of Figure~\ref{fig:VaR-256-box} for $n=256$.}
\label{fig:VaR-256}
\end{figure}

With \code{mayplot()} it is easy to create a graphical result (a pdf file for
inclusion in a \LaTeX\ document, for example)\footnote{Note that we use the
  system tool \code{pdfcrop} to crop the graph after it is generated. This
  allows one to perfectly align the graph in a
  \LaTeX\ (\file{.tex}) or Sweave (\file{.Rnw}) document.}.
Figures~\ref{fig:VaR-256-box} and \ref{fig:VaR-256} display the results for
$n=256$.
The former shows boxplots of all the $N_{sim}$ simulated Value-at-Risk estimates
$\widehat{\VaR}_\alpha(L)$, whereas the latter depicts corresponding robust Huber
  ``means'' and also demonstrates \code{mayplot()} for $N_{sim}=1$ or, equivalently, no $N_{sim}$ at all.
Overall, we see that a graphic such as Figure~\ref{fig:VaR-256-box} is
easier to grasp and to infer conclusions from than Table~\ref{tab:ft}.

% Main plot 1
%% ## log="y" makes more sense, but is less revealing to the uninitiated
%% --> in the first plot, *no* log scale (in the second, yes!)
\setkeys{Gin}{width=\textwidth}

% Main plot 2

% and this is invisible in the paper, but *is* tested (notably interactively):

%%% Local Variables:
%%% TeX-master: "parallel.tex"
%%% End:

\clearpage%%<---- Allow to print 2nd part separately !

% patchDVI setup
\Sconcordance{concordance:05_behind.tex:05_behind.Rnw:%
1 8 1 1 5 219 1 1 4 3 0 1 5 3 0 1 3 1 0 1 2 1 0 1 3 1 0 4 1 5 %
0 1 1 1 3 8 0 1 1 5 0 2 1 6 0 1 13 3 1}

% Sweave options
% use pdfcrop to crop .pdf
% put figures in ./ and name them with prefix "fig-"
%\VignetteDepends{simsalapar}
%\VignetteDepends{doSNOW}
%\VignetteDepends{foreach}
%\VignetteDepends{copula}

\section[Behind the scenes: Features of simsalapar]{%
  Behind the scenes: Advanced features of \pkg{simsalapar}}\label{sec:behind}

\subsection{Select functions for conducting the simulation}\label{sec:select:sim}

\subsubsection[The function doCallWE()]{The function \code{doCallWE()}}
The \R\ package \pkg{simsalapar} provides the following auxiliary function
\code{doCallWE()} for computing the components \code{value}, \code{error},
\code{warning}, and \code{time} as addressed in Section~\ref{sec:res}.
It is called from \code{subjob()} and based on \code{tryCatch.W.E()} which is
part of \R's \code{demo(error.catching)} for catching both warnings and errors.
\begin{lstlisting}[style=input, style=Rstyle, linerange=doCallWE-end, firstnumber=1]
## Copyright (C) 2012 Marius Hofert and Martin Maechler
##
## This program is free software; you can redistribute it and/or modify it under
## the terms of the GNU General Public License as published by the Free Software
## Foundation; either version 3 of the License, or (at your option) any later
## version.
##
## This program is distributed in the hope that it will be useful, but WITHOUT
## ANY WARRANTY; without even the implied warranty of MERCHANTABILITY or FITNESS
## FOR A PARTICULAR PURPOSE. See the GNU General Public License for more
## details.
##
## You should have received a copy of the GNU General Public License along with
## this program; if not, see <http://www.gnu.org/licenses/>.


mkTimer <- function(gcFirst) {
    if(gcFirst)
	 function(expr) 1000 * system.time(expr, gcFirst=TRUE)[[1]]
    else function(expr) 1000 * system.time(expr, gcFirst=FALSE)[[1]]
}

##' @title Safety Wrapper for Innermost Computations
##' @param f a function, given data + parameters, computes statistic
##' @param argl list of arguments for f()
##' @param timer a function like \code{\link{system.time}()}, by default,
##'              measure user time in milliseconds
##' @return list with components:
##'	    value:   f(<argl>) if that worked, NULL otherwise
##'	    error:   error message (of class simpleError) or NULL
##'	    warning: warning message (of class simpleWarning) or NULL
##'	    time:    time measured by timer()
##' @author Marius Hofert, Martin Maechler
##' { doCallWE }
doCallWE <- function(f, argl, timer = mkTimer(gcFirst=FALSE))
{
    tim <- timer( res <- tryCatch.W.E( do.call(f, argl) )) # compute f(<argl>)
    is.err <- is(val <- res$value, "simpleError") # logical indicating an error
    list(value   = if(is.err) NULL else val, # value (or NULL in case of error)
	 error   = if(is.err) val else NULL, # error (or NULL if okay)
	 warning = res$warning, # warning (or NULL)
	 time    = tim) # time
}
##' { end }
\end{lstlisting}

\subsubsection[The function subjob()]{The function \code{subjob()}}
\code{subjob()} calls \code{doOne()} via \code{doCallWE()} for computing a
sub-job, that is, a row of the virtual grid. It is called by the \code{do*()}
functions. Besides catching errors and warnings, and measuring run time via
calling \code{doCallWE()}, the main duty of \code{subjob()} is to correctly deal
with the seed. It also provides a monitor feature.
\begin{lstlisting}[style=input, style=Rstyle, linerange=subjob-end, firstnumber=1]
## Copyright (C) 2012 Marius Hofert and Martin Maechler
##
## This program is free software; you can redistribute it and/or modify it under
## the terms of the GNU General Public License as published by the Free Software
## Foundation; either version 3 of the License, or (at your option) any later
## version.
##
## This program is distributed in the hope that it will be useful, but WITHOUT
## ANY WARRANTY; without even the implied warranty of MERCHANTABILITY or FITNESS
## FOR A PARTICULAR PURPOSE. See the GNU General Public License for more
## details.
##
## You should have received a copy of the GNU General Public License along with
## this program; if not, see <http://www.gnu.org/licenses/>.


## error message for invalid seed
.invalid.seed.msg <-
    "invalid 'seed' [NULL / integer(n.sim) / list / NA / \"seq\"]"

##' @title Create a List of Advanced .Random.seed's for "L'Ecuyer-CMRG"
##' @param n number of steps to advance .Random.seed
##' @return a list of length n containing the advanced .Random.seed's
##' @author Marius Hofert and Martin Maechler
LEseeds <- function(n) {
    stopifnot(identical(RNGkind()[1], "L'Ecuyer-CMRG"), n >= 1)
    r <- vector("list", n)
    r[[1]] <- .Random.seed
    for(i in seq_len(n-1))
	r[[i+1]] <- nextRNGStream(r[[i]]) # from parallel
    r
}

## printInfo
printInfo <- local({

    strv <- function(val)
	paste("value:",
              sub("^ *", " ",capture.output(str(unname(val)))))

    SysI <- function()
        paste(format(Sys.time(), "%H:%M:%S"), "on", Sys.info()[["nodename"]])

    F <- function(cc) format(cc, justify = "right")

    formG <- function(gr, j) {
	paste(names(gr), as.matrix(F(gr))[j,], sep= "=", collapse= ", ")
    }

    ##' @title Print Diagnostics for subjob()
    ##' @param i.sim integer running in 1:n.sim
    ##' @param j integer giving the row of the physical grid pGrid
    ##'        (i.sim and j together determine the subjob() argument i)
    ##' @param pGrid physical grid; see subjob()
    ##' @param res4 4-list containing value, error, warning, and time
    ##' @param n.sim: "the" n.sim, possibly NULL
    ##' @return string with information about the sub-job just finished
    ##' @author Marius Hofert and Martin Maechler
    defMoni <- function(i.sim, j, pGrid, res4, n.sim, file="") {
	cat(SysI(), if(length(n.sim) && n.sim > 1)
	    sprintf(": i.sim=%*d, ", 1+floor(log10(n.sim)), i.sim) else ": ",
	    formG(pGrid, j), "; ", strv(res4$value), "\n",
	    sep="", file=file, append=(file != ""))
    }

    list(default = defMoni,
         gfile = function(...) defMoni(..., file="subjob-monitor.txt"),
         fileEach = function(i.sim, j, ...) {
             defMoni(i.sim, j, ..., file=sprintf("subjob-mon-%04d.txt", j))
         })
})

##' @title Function for Computing one Row of the Virtual Grid (subjob)
##' @param i row number of the virtual grid
##' @param pGrid (physical) grid with all combinations of variables of type "grid" as
##'        returned by mkGrid()
##' @param nonGrids values of non-"grid"-variables (if provided, passed to doOne())
##' @param n.sim number of simulation replications
##' @param seed one of:
##'     NULL: .Random.seed remains untouched; if it doesn't exist, generate
##'           it by calling runif(1); non-reproducible
##'     numeric(n.sim): seeds (numbers) for each of the n.sim simulation
##'                     replications (same seed for each row in the (physical)
##'                     grid); reproducible
##'     vector("list", n.sim): seeds (vectors) for each of the n.sim simulation
##'                            replications (same seed for each row in the (physical)
##'                            grid); this case is meant for the rng L'Ecuyer-CMRG;
##'                            reproducible
##'     NA: .Random.seed remains untouched; if it doesn't exist, so be it;
##'         no 5th component is concatenated to the result of the doOne() call;
##'         non-reproducible
##'     character string: specifying a seeding method; currently only "seq" for
##'                       the seeds 1:n.sim for the n.sim simulation replications;
##'                       reproducible
##' @param keepSeed logical indicating whether seed is 'stored'/appended
##' @param repFirst logical; if FALSE (the default), all n.sim replications are computed for
##'             a specific row in the (physical) grid before the next row is considered;
##'             if TRUE, first all rows of the (physical) grid are computed for a fixed
##'             replication until the next replication is considered.
##' @param doOne function for computing one row in the (physical) grid; must return
##'        the return value of doCallWE()
##' @param monitor logical or function, see help file
##' @param ... additional arguments passed to doOne()
##' @return a vector of length 5 if seed!=NA, otherwise of length 4. The first four
##'         components contain the return value of doOne() which should be the return
##'         value of doCallWE(). The 5th component contains .Random.seed before the
##'         call of doCallWE() (for reproducibility).
##' @note subjob() is an auxiliary function called from doLapply(), doMclapply(),
##'       doClusterApply() etc.
##' @author Marius Hofert and Martin Maechler
##' { subjob }
subjob <- function(i, pGrid, nonGrids, n.sim, seed, keepSeed=FALSE,
                   repFirst=TRUE, doOne,
                   timer=mkTimer(gcFirst=FALSE), monitor=FALSE, ...)
{
    ## i |-> (i.sim, j) :
    ## determine corresponding i.sim and row j in the physical grid
    if(repFirst) {
	i.sim <- 1 + (i-1) %%  n.sim ## == i  when n.sim == 1
	j     <- 1 + (i-1) %/% n.sim ## row of pGrid
	## Note: this case first iterates over i.sim, then over j:
	## (i.sim,j) = (1,1), (2,1), (3,1),..., (1,2), (2,2), (3,2), ...
    } else {
	ngr <- nrow(pGrid) # number of rows of the (physical) grid
	j     <- 1 + (i-1) %%  ngr ## row of pGrid
	i.sim <- 1 + (i-1) %/% ngr
	## Note: this case first iterates over j, then over i.sim:
	## (i.sim,j) = (1,1), (1,2), (1,2),..., (2,1), (2,2), (2,3), ...
    }

    ## seeding
    if(is.null(seed)) {
	if(!exists(".Random.seed")) runif(1) # guarantees that .Random.seed exists
	## => this is typically not reproducible
    }
    else if(is.numeric(seed)) {
	if(length(seed) != n.sim) stop("'seed' has to be of length ", n.sim)
	set.seed(seed[i.sim]) # same seed for all runs within the same i.sim
	## => calculations based on same random numbers as much as possible
    }
    ## else if(length(seed) == n.sim*ngr && is.numeric(seed)) {
    ##    set.seed(seed[i]) # different seed for *every* row of the virtual grid
    ##    always (?) suboptimal (more variance than necessary)
    ## }
    else if(is.list(seed)) { # (currently) L'Ecuyer-CMRG
	if(length(seed) != n.sim) stop("'seed' has to be of length ", n.sim)
	if(!exists(".Random.seed"))
	    stop(".Random.seed does not exist - in l'Ecuyer setting")
	assign(".Random.seed", seed[[i.sim]], envir = globalenv())
    }
    else if(is.na(seed)) {
	keepSeed <- FALSE
    }
    else {
	if(!is.character(seed)) stop(.invalid.seed.msg)
	switch(match.arg(seed, choices = c("seq")),
	       "seq" = { # sequential seed :
		   set.seed(i.sim) #same seed for all runs within the same i.sim
		   ## => calculations based on the same random numbers
	       },
	       stop("invalid character 'seed': ", seed)
	       )
    }
    ## save seed, compute and return result for one row of the virtual grid
    if(keepSeed) rs <- .Random.seed # <- save here in case it is advanced in doOne

    ## monitor checks happen already in caller!
    if(isTRUE(monitor)) monitor <- printInfo[["default"]]

    ## doOne()'s arguments, grids, non-grids, and '...':
    args <- c(pGrid[j, , drop=FALSE],
	      ## [nonGrids is never missing when called from doLapply() etc.]
	      if(missing(nonGrids) || length(nonGrids) == 0)
	      list(...) else c(nonGrids, ...))
    nmOne <- names(formals(doOne))
    if(!identical(nmOne, "..."))
	args <- args[match(names(args), nmOne)] # adjust order for doOne()

    r4 <- doCallWE(doOne, args, timer = timer)

    ## monitor (after computation)
    if(is.function(monitor)) monitor(i.sim, j=j, pGrid=pGrid, n.sim=n.sim, res4=r4)

    c(r4, if(keepSeed) list(.Random.seed = rs)) # 5th component .Random.seed
}
##' { end }

##' Check a user's "doOne" function; do similar things as mkAL():
doCheck <- function(doOne, vList, nChks = ng, verbose=TRUE)
{
    stopifnot(is(vList, "varlist"))
    pGrid <- mkGrid(vList)
    nonGrids <- get.nonGrids(vList)$nonGrids
    stopifnot(is.function(doOne), is.data.frame(pGrid), is.list(nonGrids),
	      is.integer(ng <- nrow(pGrid)), ng >= 1)
    j.s <- sample.int(ng, nChks)
    for(j in j.s) {
	if(verbose) cat(sprintf("j =%3d:", j))
	## doOne()'s arguments, grids, non-grids, and '...':
	gr <- pGrid[j, , drop=FALSE]
	args <- if(length(nonGrids) == 0) as.list(gr) else c(gr, nonGrids)
	nmOne <- names(formals(doOne))
	if(!identical(nmOne, "..."))# fix order for doOne()
	    args <- args[match(names(args), nmOne)]

	r <- do.call(doOne, args)
	dr <- dim(r)
	if(verbose) {
	    cat(" -> length(r) = ",length(r)," dim(r):"); print(dr)
	    cat("str(dimnames(r)):"); str(dimnames(r))
        }
	stopifnot(is.numeric(r))
        if(is.null(dr)) dr <- length(r)
	## TODO MM: check "inner" --map--> dim & dimnames of r
        ## -------  in the same way as mkAL()
        dnr <- dimnames(r)
        ni <- length(l.i <- sapply(getEl(vList, "inner"), length))
        if(anyDuplicated(dr) && is.null(names(dnr)))
            warning("some dimensions in your doOne() result are equal\n",
                    "and cannot be distinguished by names(dimnames(.))")
        ## TODO MM: check names(dnr) *iff* mkAL() uses the same, i.e. not yet
        if(!all(tail(dr, ni) == l.i))
	    stop(gettextf("tail(dim(doOne(..)), %d) differs from %s",
			  ni, 'sapply(getEl(vList, "inner"), length)'),
                 domain=NA)
    }
}


\end{lstlisting}

The different seeding methods implemented are:
\begin{itemize}
\item \code{NULL}: In this case \code{.Random.seed} remains
  untouched. If it does not exist, it is generated by calling
  \code{runif(1)}. With this seeding method, the results are typically not reproducible.
\item A \code{numeric} vector, say \code{s}, of length \code{n.sim}, providing
  seeds for each of the \code{n.sim} simulation replications, i.e., simulation
  \code{i} receives seed \code{set.seed(s[i])}, for \code{i} from 1 to
  \code{n.sim}. For a fixed replication \code{i}, the seed is the same no matter
  what row in the (physical) grid is considered. This ensures least variance
  across the computations for the same replication \code{i}. In particular, it
  also leads to the same results no matter which variables are of type ``grid''
  or ``inner''; see \code{demo(robust.mean)} where this is tested. This is
  important to guarantee since one might want to change certain ``inner''
  variables to ``grid'' variables due to load-balancing while computing the
  desired statistics based on the same seed (or generated data from this
  seed). Clearly, since replication \code{i} is guaranteed to get seed
  \code{s[i]} (no matter when the corresponding sub-job is computed relative to
  all other sub-jobs), this seeding method provides reproducible results.
\item A \code{list} of length \code{n.sim} which provides seeds for
  each of the \code{n.sim} simulation replications. In contrast to the case
  of a \code{numeric} vector, this case is meant to be for providing more
  general seeds. At the moment, seeds for l'Ecuyer's random number generator
  \code{L'Ecuyer-CMRG} can be provided; see \cite{lecuyersimardchenkelton2002}
  for a reference and Section~\ref{sec:behind:varlists} for how to use it. This seeding method also provides
  reproducible results.
\item \code{NA}: In this case \code{.Random.seed} remains untouched. In
  contrast to \code{NULL}, it is not even generated if it does not exist. Also,
  the fifth component \code{.Random.seed} is not concatenated to the result in
  this case. In all other cases, it is appended if \code{keepSeed=TRUE}. As
  mentioned before, the default \code{keepSeed=FALSE} has been chosen
  to avoid large result objects. Clearly, seeding method \code{NA} typically does not
  provide reproducible results.
\item a \code{character} string, specifying a certain seeding method. Currently,
  only \code{"seq"} is provided, a convenient special case of the second case
  addressed above, where the vector of seeds is simply \code{1:n.sim}, and thus provides
  reproducible results.
\end{itemize}
If \code{keepSeed=TRUE} and \code{seed} is not \code{NA}, \code{subjob()}
saves \code{.Random.seed} as the fifth component of the output vector
(besides the four components returned by \code{doCallWE()}). This is useful for
reproducing the result of the corresponding call of \code{doOne()} for debugging
purposes, for example.

The default seeding method in the \code{do*()} functions is \code{"seq"}. This
is a comparably simple default which guarantees reproducibility. Note, however,
that for very large simulations, there is no guarantee that the random-number
streams are sufficiently ``apart''. For this, we recommend l'Ecuyer's random
number generator \code{L'Ecuyer-CMRG}; see Section~\ref{sec:behind:varlists} for an example.

\subsubsection[The function doLapply()]{The function \code{doLapply()}}
As mentioned before, \code{doLapply()} is essentially a wrapper for
\code{lapply()} to iterate (sequentially) over all rows in the virtual grid,
that is, over all sub-jobs. As an important ingredient, \code{saveSim()}, explained below, is used
to deal with the raw result list.
\begin{lstlisting}[style=input, style=Rstyle, linerange=doLapply-end, firstnumber=1]
## Copyright (C) 2012-13 Marius Hofert and Martin Maechler
##
## This program is free software; you can redistribute it and/or modify it under
## the terms of the GNU General Public License as published by the Free Software
## Foundation; either version 3 of the License, or (at your option) any later
## version.
##
## This program is distributed in the hope that it will be useful, but WITHOUT
## ANY WARRANTY; without even the implied warranty of MERCHANTABILITY or FITNESS
## FOR A PARTICULAR PURPOSE. See the GNU General Public License for more
## details.
##
## You should have received a copy of the GNU General Public License along with
## this program; if not, see <http://www.gnu.org/licenses/>.

##' @title Function for Iterating Over All Subjobs (Non-Parallel)
##' @param vList list of variable specifications
##' @param seed repFirst: see subjob()
##' @param repFirst see subjob()
##' @param sfile see saveSim()
##' @param check see saveSim()
##' @param doAL see saveSim()
##' @param subjob. function for computing a subjob (one row of the virtual grid);
##'        typically subjob()
##' @param doOne user-supplied function for computing one row of the (physical)
##'        grid
##' @param ... additional arguments passed to subjob() (typically further
##'        passed on to doOne())
##' @return the result of applying subjob() to all subjobs, converted with saveSim()
##' @author Marius Hofert and Martin Maechler
##' @note Works *sequentially*
##' { doLapply }
doLapply <- function(vList, seed="seq", repFirst=TRUE, sfile=NULL,
                     check=TRUE, doAL=TRUE, subjob.=subjob, monitor=FALSE,
                     doOne, ...)
{
    if(!is.null(r <- maybeRead(sfile))) return(r)
    stopifnot(is.function(subjob.), is.function(doOne))
    if(!(is.null(seed) || is.na(seed) || is.numeric(seed) ||
         (is.list(seed) && all(vapply(seed, is.numeric, NA))) ||
         is.character(seed) ))
        stop(.invalid.seed.msg)
    if(check) doCheck(doOne, vList, nChks=1, verbose=FALSE)

    ## monitor checks {here, not in subjob()!}
    if(!(is.logical(monitor) || is.function(monitor)))
        stop(gettextf("'monitor' must be logical or a function like %s",
                      'printInfo[["default"]]'))

    ## variables
    pGrid <- mkGrid(vList)
    ngr <- nrow(pGrid)
    ng <- get.nonGrids(vList) # => n.sim >= 1
    n.sim <- ng$n.sim # get n.sim

    ## actual work
    res <- lapply(seq_len(ngr * n.sim), subjob.,
                  pGrid=pGrid, nonGrids = ng$nonGrids, repFirst=repFirst,
                  n.sim=n.sim, seed=seed, doOne=doOne, monitor=monitor, ...)

    ## convert result and save
    saveSim(res, vList=vList, repFirst=repFirst,sfile=sfile,check=check,doAL=doAL)
}
##' { end } doLapply

##' @title Function for Iterating Over All Subjobs in Parallel Using Foreach
##' @param vList list of variable specifications
##' @param doCluster logical indicating whether the sub jobs are run on a cluster
##'        or rather several cores
##' @param spec if doCluster=TRUE : number of nodes; passed to parallel's
##'                                 makeCluster()
##'             if doCluster=FALSE: number of cores
##' @param type cluster type, see parallel's ?makeCluster
##' @param block.size size of blocks of rows in the virtual grid which are computed
##'        simultaneously
##' @param seed see subjob()
##' @param repFirst see subjob()
##' @param sfile see saveSim()
##' @param check see saveSim()
##' @param doAL see saveSim()
##' @param subjob. function for computing a subjob (one row of the virtual grid);
##'        typically subjob()
##' @param doOne user-supplied function for computing one row of the (physical)
##'        grid
##' @param extraPkgs character vector of packages to be made available on the nodes
##' @param exports character vector of functions to export
##' @param ... additional arguments passed to subjob() (typically further
##'        passed on to doOne())
##' @return result of applying subjob() to all subjobs, converted with saveSim()
##' @author Marius Hofert and Martin Maechler
##' @note Works on multiple nodes or cores
##' { doForeach }
doForeach <- function(vList, doCluster = !(missing(spec) && missing(type)),
                      spec=detectCores(), type="MPI", block.size=1,
                      seed="seq", repFirst=TRUE,
                      sfile=NULL, check=TRUE, doAL=TRUE,
                      subjob.=subjob, monitor=FALSE, doOne,
                      extraPkgs=character(), exports=character(), ...)
{
    ## Unfortunately, imports() ends not finding 'iter' from pkg "iterators":
    ## --> rather strictly require things here:
    stopifnot(require("foreach"), require("doParallel"))
    if(!is.null(r <- maybeRead(sfile))) return(r)
    stopifnot(is.function(subjob.), is.function(doOne))
    if(!(is.null(seed) || is.na(seed) || is.numeric(seed) ||
         (is.list(seed) && all(vapply(seed, is.numeric, NA))) ||
         is.character(seed) ))
        stop(.invalid.seed.msg)
    if(check) doCheck(doOne, vList, nChks=1, verbose=FALSE)

    ## monitor checks {here, not in subjob()!}
    if(!(is.logical(monitor) || is.function(monitor)))
        stop(gettextf("'monitor' must be logical or a function like %s",
                      'printInfo[["default"]]'))

    ## variables
    pGrid <- mkGrid(vList)
    ngr <- nrow(pGrid)
    ng <- get.nonGrids(vList) # => n.sim >= 1
    n.sim <- ng$n.sim
    stopifnot(1 <= block.size, block.size <= n.sim, n.sim %% block.size == 0)

    ## Two main cases for parallel computing
    if(!doCluster) { # multiple cores
        ## ?registerDoParallel -> Details -> Unix + multiple cores => 'fork' is used
        stopifnot(is.numeric(spec), length(spec) == 1)
        registerDoParallel(cores=spec) # register doParallel to be used with foreach
    }
    else { # multiple nodes
        ## One actually only needs makeCluster() when setting up a *cluster*
        ## for working on different nodes. In this case, the 'spec' argument
        ## specifies the number of nodes.
        ## The docu about registerDoParallel() might be slightly misleading...
        cl <- makeCluster(spec, type=type) # create cluster
        on.exit(stopCluster(cl)) # shut down cluster and execution environment
        registerDoParallel(cl)  # register doParallel to be used with foreach
    }
    if(check) cat(sprintf("getDoParWorkers(): %d\n", getDoParWorkers()))

    ## actual work
    n.block <- n.sim %/% block.size
    i <- NULL ## <- dirty but required for R CMD check ...
    res <- ul(foreach(i=seq_len(ngr * n.block),
                      .packages=c("simsalapar", extraPkgs),
                      .export=c(".Random.seed", "iter", "mkTimer", exports)) %dopar%
          {
              lapply(seq_len(block.size), function(k)
                     subjob.((i-1)*block.size+k, pGrid=pGrid,
                             nonGrids=ng$nonGrids, repFirst=repFirst,
                             n.sim=n.sim, seed=seed, doOne=doOne,
                             monitor=monitor, ...))})
    ## convert result and save
    saveSim(res, vList, repFirst=repFirst, sfile=sfile, check=check, doAL=doAL)
}
##' { end } doForeach

##' @title Function for Iterating Over All Subjobs in Parallel Using Rmpi
##' @param vList list of variable specifications
##' @param spec cluster specification (number of workers)
##' @param load.balancing logical indicating whether to use mpi.applyLB() instead of mpi.apply()
##' @param block.size size of blocks of rows in the virtual grid which are computed
##'        simultaneously
##' @param seed see subjob()
##' @param repFirst see subjob()
##' @param sfile see saveSim()
##' @param check see saveSim()
##' @param doAL see saveSim()
##' @param subjob. function for computing a subjob (one row of the virtual grid);
##'        typically subjob()
##' @param doOne user-supplied function for computing one row of the (physical)
##'        grid
##' @param exports vector of objects to export
##' @param ... additional arguments passed to subjob() (typically further
##'        passed on to doOne())
##' @return the result of applying subjob() to all subjobs, converted with saveSim()
##' @author Marius Hofert and Martin Maechler
##' @note Works on multiple nodes or cores
##' Email from Rmpi maintainer (hyu@stats.uwo.ca) on 2013-06-10:
##' If you are using OpenMPI, then mpi.universe.size() will always return 1
##' unless R is launched through mpirun.
##' Yes. You can use the option nslaves to launch slaves as many as you want.
##' How those slave processes assigned to nodes/cores are  controlled by
##' OpenMPI (different MPIs have different ways of assigning slave processes
##' but most recycle available notes/cores). In your cases, you probably
##' choose nslaves=4 so that all cores are running in parallel. However,
##' setting nslaves to be higher than the available notes/codes achieves some
##' kind loading balancing. For example, nslaves = 8 essentially spreads an
##' entire job into 8 small ones instead of 4 small ones. This gives some
##' advantages if one of the original 4 small jobs runs much longer than
##' others.
##' mpi.universe.size() # => 1; the total number of CPUs available in a cluster
##' mpi.spawn.Rslaves() # spawn as many slaves as the MPI environment knows (=> 1 master, 1 slave)
##' mpi.close.Rslaves()
##' mpi.spawn.Rslaves(nslaves=17) # spawn more slaves than possible (?) (=> 1
##'                                 master, 17 slaves) => calculations are still
##'                                 only done on the max. available cores; see
##'                                 test script http://collaborate.bu.edu/linga/ParallelMCMC
##' mpi.close.Rslaves()
##' Note: spawning more slaves than available may lead to errors (MH)
##' { doRmpi }
doRmpi <- function(vList,
                   nslaves = if((sz <- mpi.universe.size()) <= 1) detectCores()
                             else sz,
                   load.balancing=TRUE, block.size=1, seed="seq", repFirst=TRUE,
                   sfile=NULL, check=TRUE, doAL=TRUE, subjob.=subjob, monitor=FALSE,
                   doOne, exports=character(), ...)
{
    if(!require("Rmpi"))
        stop("You must install the CRAN package 'Rmpi' before you can use doRmpi()")

    if(!is.null(r <- maybeRead(sfile))) return(r)
    stopifnot(is.function(subjob.), is.function(doOne))
    if(!(is.null(seed) || is.na(seed) || is.numeric(seed) ||
         (is.list(seed) && all(vapply(seed, is.numeric, NA))) ||
         is.character(seed) ))
        stop(.invalid.seed.msg)
    if(check) doCheck(doOne, vList, nChks=1, verbose=FALSE)

    ## monitor checks {here, not in subjob()!}
    if(!(is.logical(monitor) || is.function(monitor)))
        stop(gettextf("'monitor' must be logical or a function like %s",
                      'printInfo[["default"]]'))

    ## variables
    pGrid <- mkGrid(vList)
    ngr <- nrow(pGrid)
    ng <- get.nonGrids(vList) # => n.sim >= 1
    n.sim <- ng$n.sim
    stopifnot(1 <= block.size, block.size <= n.sim, n.sim %% block.size == 0)

    ## use as many workers as available
    ## Note: mpi.comm.size(comm) returns the total number of members in a comm
    comm <- 1  ## communicator number
    if (!mpi.comm.size(comm)) ## <==> no slaves are running
        mpi.spawn.Rslaves(nslaves=nslaves)
    ## quiet = TRUE would omit successfully spawned slaves
    on.exit(mpi.close.Rslaves()) # close slaves spawned by mpi.spawn.Rslaves()
    ## pass global required objects to cluster (required by mpi.apply())
    mpi.bcast.Robj2slave(.Random.seed)
    mpi.bcast.Robj2slave(mkTimer)
    for(e in exports) {
        ee <- substitute(mpi.bcast.Robj2slave(EXP), list(EXP = as.symbol(e)))
        eval(ee)
    }

    ## instead of initExpr, this needs a 'initFunction' + 'initArgs'
    ## if(!missing(initExpr)) do.call(mpi.bcast.cmd, c(list(initFunction), ...))

    ## actual work
    n.block <- n.sim %/% block.size
    res <- ul((if(load.balancing) mpi.applyLB else mpi.apply)(
        seq_len(ngr * n.block), function(i)
        lapply(seq_len(block.size), function(k)
            subjob.((i-1)*block.size+k, pGrid=pGrid,
                    nonGrids=ng$nonGrids, repFirst=repFirst,
                    n.sim=n.sim, seed=seed, doOne=doOne, monitor=monitor, ...))))

    ## convert result and save
    saveSim(res, vList, repFirst=repFirst, sfile=sfile, check=check, doAL=doAL)
}
##' { end } doRmpi

##' @title Function for Iterating Over All Subjobs in Parallel Using mclapply()
##' @param vList list of variable specifications
##' @param cores number of cores
##' @param load.balancing logical indicating whether to use mpi.applyLB() instead of mpi.apply()
##' @param block.size size of blocks of rows in the virtual grid which are computed
##'        simultaneously
##' @param seed see subjob()
##' @param repFirst see subjob()
##' @param sfile see saveSim()
##' @param check see saveSim()
##' @param doAL see saveSim()
##' @param subjob. function for computing a subjob (one row of the virtual grid);
##'        typically subjob()
##' @param doOne user-supplied function for computing one row of the (physical)
##'        grid
##' @param ... additional arguments passed to subjob() (typically further
##'        passed on to doOne())
##' @return the result of applying subjob() to all subjobs, converted with saveSim()
##' @author Marius Hofert and Martin Maechler
##' @note Works on multiple cores (but runs *sequentially* on Windows)
##' { doMclapply }
doMclapply <-
    function(vList,
             cores = if(.Platform$OS.type == "windows") 1 else detectCores(),
             load.balancing=TRUE, block.size=1, seed="seq", repFirst=TRUE,
             sfile=NULL, check=TRUE, doAL=TRUE, subjob.=subjob,
             monitor=FALSE, doOne, ...)
{
    if(!is.null(r <- maybeRead(sfile))) return(r)
    stopifnot(is.function(subjob.), is.function(doOne))
    if(!(is.null(seed) || is.na(seed) || is.numeric(seed) ||
         (is.list(seed) && all(vapply(seed, is.numeric, NA))) ||
         is.character(seed) ))
        stop(.invalid.seed.msg)
    if(check) doCheck(doOne, vList, nChks=1, verbose=FALSE)

    ## variables
    pGrid <- mkGrid(vList)
    ngr <- nrow(pGrid)
    ng <- get.nonGrids(vList) # => n.sim >= 1
    n.sim <- ng$n.sim
    stopifnot(1 <= block.size, block.size <= n.sim, n.sim %% block.size == 0)

    ## monitor checks
    if(!(is.logical(monitor) || is.function(monitor)))
        stop(gettextf("'monitor' must be logical or a function like %s",
                      'printInfo[["default"]]'))

    ## actual work
    n.block <- n.sim %/% block.size
    res <- ul(mclapply(seq_len(ngr * n.block), function(i)
                       lapply(seq_len(block.size), function(k)
                              subjob.((i-1)*block.size+k, pGrid=pGrid,
                                      nonGrids=ng$nonGrids, repFirst=repFirst,
                                      n.sim=n.sim, seed=seed, doOne=doOne,
                                      monitor=monitor, ...)),
                       mc.cores = cores,
                       mc.preschedule = !load.balancing, mc.set.seed=FALSE))

    ## convert result and save
    saveSim(res, vList, repFirst=repFirst, sfile=sfile, check=check, doAL=doAL)
}
##' { end } doMclapply

##' @title Function for Iterating Over All Subjobs in Parallel Using clusterApply()
##' @param vList list of variable specifications
##' @param spec cluster specification (number of workers)
##' @param type cluster type, see parallel's ?makeCluster (basically snow's makeCluster)
##' @param load.balancing logical indicating whether to use clusterApplyLB()
##'     instead of clusterApply()
##' @param block.size size of blocks of rows in the virtual grid which are computed
##'        simultaneously
##' @param seed see subjob()
##' @param repFirst see subjob()
##' @param sfile see saveSim()
##' @param check see saveSim()
##' @param doAL see saveSim()
##' @param subjob. function for computing a subjob (one row of the virtual grid);
##'        typically subjob()
##' @param doOne user-supplied function for computing one row of the (physical)
##'        grid
##' @param initExpr expression initially evaluated on the cluster (can be missing)
##' @param ... additional arguments passed to subjob() (typically further
##'        passed on to doOne())
##' @return the result of applying subjob() to all subjobs, converted with saveSim()
##' @author Marius Hofert and Martin Maechler
##' @note Works on multiple nodes or cores
##' { doClusterApply }
doClusterApply <-
    function(vList, spec=detectCores(), type="MPI",
             load.balancing=TRUE, block.size=1, seed="seq", repFirst=TRUE,
             sfile=NULL, check=TRUE, doAL=TRUE, subjob.=subjob, monitor=FALSE,
             doOne, initExpr, exports=character(), ...)
{
    if(!is.null(r <- maybeRead(sfile))) return(r)
    stopifnot(is.function(subjob.), is.function(doOne))
    if(!(is.null(seed) || is.na(seed) || is.numeric(seed) ||
         (is.list(seed) && all(vapply(seed, is.numeric, NA))) ||
         is.character(seed) ))
        stop(.invalid.seed.msg)
    if(check) doCheck(doOne, vList, nChks=1, verbose=FALSE)

    ## variables
    pGrid <- mkGrid(vList)
    ngr <- nrow(pGrid)
    ng <- get.nonGrids(vList) # => n.sim >= 1
    n.sim <- ng$n.sim
    stopifnot(1 <= block.size, block.size <= n.sim, n.sim %% block.size == 0)

    ## create cluster object
    cl <- makeCluster(spec, type=type)
    on.exit(stopCluster(cl)) ## shut down cluster and execution environment

    ## monitor checks
    if(!(is.logical(monitor) || is.function(monitor)))
        stop(gettextf("'monitor' must be logical or a function like %s",
                      'printInfo[["default"]]'))

    clusterExport(cl, varlist=c(".Random.seed", "mkTimer", exports))
    if(!missing(initExpr)) clusterCall(cl, eval, substitute(initExpr))

    ## actual work
    n.block <- n.sim %/% block.size
    res <- ul((if(load.balancing) clusterApplyLB else clusterApply)(
        cl, seq_len(ngr * n.block), function(i)
        lapply(seq_len(block.size), function(k)
               subjob.((i-1)*block.size+k, pGrid=pGrid,
                       nonGrids=ng$nonGrids, repFirst=repFirst,
                       n.sim=n.sim, seed=seed, doOne=doOne, monitor=monitor, ...))))

    ## convert result and save
    saveSim(res, vList, repFirst=repFirst, sfile=sfile, check=check, doAL=doAL)
}
##' { end } doClusterApply

##' Function for comparing do*Apply() results:
##' { doRes.equal }
doRes.equal <- function(x,y, tol=1e-15, ...)
    all.equal(lapply(x, `[`, 1:3),
              lapply(y, `[`, 1:3), tol=tol, ...)
##' { end }

\end{lstlisting}

\subsubsection[The functions saveSim() and maybeRead()]{The functions \code{saveSim()} and \code{maybeRead()}}
After having conducted the main simulation with one of the \code{do*()}
functions, we would like to create and store the result array. It can then be
loaded and worked on for the analysis of the study which is often done on a different
computer. For creating, checking, and saving the array,
\pkg{simsalapar} provides the function \code{saveSim()}.

%% \paragraph{saveSim()}
%%            =========
If possible, \code{saveSim()} creates an array of lists (via \code{mkAL()}),
where each element of the array is a list of length four or five as returned by
\code{subjob()}. If this fails, \code{saveSim()} simply takes its input list. It
then stores this array (or list) in the given \cmd{.rds} file (via \code{saveRDS()}) and
returns it for further usage. In our working example, the array itself is
five-dimensional, the dimensions corresponding to $n$, $d$, $C$, $\tau$, and
$N_{sim}$.
\begin{lstlisting}[style=input, style=Rstyle, linerange=saveSim-end, firstnumber=1]
## Copyright (C) 2012 Marius Hofert and Martin Maechler
##
## This program is free software; you can redistribute it and/or modify it under
## the terms of the GNU General Public License as published by the Free Software
## Foundation; either version 3 of the License, or (at your option) any later
## version.
##
## This program is distributed in the hope that it will be useful, but WITHOUT
## ANY WARRANTY; without even the implied warranty of MERCHANTABILITY or FITNESS
## FOR A PARTICULAR PURPOSE. See the GNU General Public License for more
## details.
##
## You should have received a copy of the GNU General Public License along with
## this program; if not, see <http://www.gnu.org/licenses/>.


##' @title Wrapper of unlist(, recursive=FALSE)
##' @param x argument
##' @return argument unlisted by one level
##' @author Marius Hofert
ul <- function(x) unlist(x, recursive=FALSE)

##' @title Converting a List to an Array of Lists
##' @param x list of length prod(dm) where each element is a list of length 5
##'	   containing the named elements "value", "error", "warning", "time", and
##'	   ".Random.seed", the first four as returned by doCallWE()
##' @param vList variable specification list
##' @param repFirst logical
##' @param check logical indicating whether checks are carried out
##' @return array of lists of length 5
##' @author Marius Hofert and Martin Maechler
##' @note Arrays are just vectors and thus have to have elements of length 1.
##'	  Therefore, the result of mkAL()  is accessed with [[,..,]],
##'	  so [[,..,]]$value gives the corresponding 'value', for example.
##' { mkAL }
mkAL <- function(x, vList, repFirst, check=TRUE)
{
    grVars <- getEl(vList, "grid", NA)
    n.sim <- get.n.sim(vList)
    ngr <- prod(vapply(lapply(grVars, `[[`, "value"), length, 1L)) # nrow(pGrid)
    lx <- n.sim * ngr
    if(check) {
	stopifnot(is.list(x))
        if(length(x) != lx)
            stop("varlist-defined grid variable dimensions do not match length(x)")
	if(length(x) >= 1) {
	    x1 <- x[[1]]
	    stopifnot(is.list(x1),
		      c("value", "error", "warning", "time") %in% names(x1))
	}
    }
    if(repFirst) ## reorder x
	x <- x[as.vector(matrix(seq_len(lx), ngr, n.sim, byrow=TRUE))]
    iVals <- getEl(vList, "inner")
    xval <- lapply(x, `[[`, "value")
    iLen <- vapply(iVals, length, 1L)
    n.inVals <- prod(iLen)
    if(check) {
	## vector of all "value" lengths
	v.len <- vapply(xval, length, 1L)
	## NB: will be of length zero, when an error occured !!

	##' is N a true multiple of D? includes equality, but we also true vector
	is.T.mult <- function(N, D) N >= D & {q <- N / D; q == as.integer(q) }

	if(!all(eq <- is.T.mult(v.len, n.inVals))) {
	    ## (!all(len.divides <- v.len %% n.inVals == 0)) {
	    not.err <- vapply(lapply(x, `[[`, "error"), is.null, NA)
	    if(!identical(eq, not.err)) {
		msg <- gettextf(
		  "some \"value\" lengths differ from 'n.inVals'=%d without error",
		  n.inVals)
		if(interactive()) {
		    ## warning() instead of stop():
		    ## had *lots* of computing till here --> want to investigate
		    warning(msg, domain=NA, immediate. = TRUE)
		    cat("You can investigate (v.len, xval, etc) now:\n")
		    browser()
		}
		else stop(msg, domain=NA)
	    }
	    if(all(v.len == 0))
		warning(gettextf(
                  "All \"%s\"s are of length zero. The first error message is\n %s",
                  "value", dQuote(conditionMessage(x[[1]][["error"]]))),
                        domain=NA)
	}
    }

    if(length(iVals) > 0 && length(xval) > 0) {
	## ensure that inner variable names are "attached" to x's "value"s :
	if(noArr <- is.null(di <- dim(xval[[1]])))
	    di <- length(xval[[1]])
	rnk <- length(di)# true dim() induced rank
	nI <- length(iLen)# = number of inner Vars;  iLen are their lengths
	for(i in seq_along(xval)) {
	    n. <- length(xi <- xval[[i]])
	    if(n. == 0) # 'if (check)' above has already ensured this is an "error"
		xi <- NA_real_
	    ## else if (n. != n.inVals)
	    ##	   warning(gettext("x[[%d]] is of wrong length (=%d) instead of %d",
	    ##			   i, n., n.inVals), domain=NA)
	    dn.i <- if(noArr) {
		if(nI == 1) list(names(xi)) else rep.int(list(NULL), nI)
	    } else if(is.null(dd <- dimnames(xi))) rep.int(list(NULL), rnk) else dd
	    ## ==>  rnk := length(di) == length(dn.i)
	    if(rnk == nI)# = length(iVals) = length(iLen)  --  simple matching case
		names(dn.i) <- names(iLen)
	    else { # more complicated as doOne() returned a full vector, matrix ...
                if(rnk != length(dn.i)) warning(
                "dim() rank, i.e., length(dim(.)), does not match dimnames() rank")
		if(nI > rnk) # or rather error?
		    warning("nI=length(iVals) larger than length(<dimnames>)")
		else { # nI<rnk==length(di)==length(dn.i) => find matching dim()
		    ## assume inner variables match the *end* of the array
		    j <- seq_len(rnk - nI)
		    j <- which(di[nI+ j] == iLen[j])
		    if(is.null(names(dn.i))) names(dn.i) <- rep.int("", rnk)
		    names(dn.i)[nI+j] <- names(iLen)[j]
		}
	    }
	    x[[i]][["value"]] <- array(xi, dim=if(noArr)iLen else di, dimnames=dn.i)
	}
    }

    gridNms <- mkNms(grVars, addNms=TRUE)
    dmn <- lapply(gridNms, sub, pattern=".*= *", replacement="")
    dm <- vapply(dmn, length, 1L)
    if(n.sim > 1) {
	dm <- c(dm, n.sim=n.sim)
	dmn <- c(dmn, list(n.sim=NULL))
    }
    ## build array
    array(x, dim=dm, dimnames=dmn)
}
##' { end }

##' @title Converting a List to an Array of Lists, Save and Return It
##' @param x result list from mclapply(), clusterApply() etc.
##' @param vList variable specification list
##' @param sfile file name with extension .rds; if sfile=NULL, then nothing is
##'	   saved (and saveSim() is equal to mkAL())
##' @param check logical indicating whether checks are carried out
##' @param doAL logical indicating whether mkAL() should be called
##' @return 'x' or the array returned by mkAL()
##' @author Martin Maechler
##' { saveSim }
saveSim <- function(x, vList, repFirst, sfile, check=TRUE, doAL=TRUE)
{
    if(doAL) {
	a <- tryCatch(mkAL(x, vList, repFirst=repFirst, check=check),
                      error=function(e) e)
	if(inherits(a, "error")) {
	  warning(paste(
            "Relax..: The simulation result 'x' is being saved;",
            "we had an error in 'mkAL(x, *)' ==> returning 'x' (argument, a list).",
            "  you can investigate mkAL(x, ..) yourself. The mkAL() err.message:",
            conditionMessage(a), sep="\n"))
	  a <- x
	}
    } else a <- x
    if(!is.null(sfile))
	saveRDS(a, file=sfile)
    a
}
##' { end }

##' @title Possibly Read Object from an .rds File
##' @param sfile file name with extension .rds
##' @param msg logical indicating whether a message is printed when an object is read
##' @return the object, or NULL (if the file does not exist)
##' @author Martin Maechler
##' { maybeRead }
maybeRead <- function(sfile, msg=TRUE)
{
    if(is.character(sfile) && file.exists(sfile)) {
	if(msg) message("getting object from ", sfile)
	structure(readRDS(sfile), fromFile = TRUE)
    }
}
##' { end }

##' @title Compute Array of Simulation Result Values
##' @param x typically, resulting from mkAL(), an array of 5-lists with
##'	   components "value", "error", "warning", "time", and ".Random.seed".
##' @param err.value numeric value which is used in case of an error
##' @param FUN function to be applied before the array is built
##' @return array of values or err.value (in case of an error)
##' @author Marius Hofert and Martin Maechler
valArray <- function(x, err.value = NA, FUN = NULL)
{
    dmn <- dimnames(x)
    dm <- dim(x)
    rr <- lapply(x, `[[`, "value")
    no.err <- sapply(lapply(x, `[[`, "error"), is.null)
    r <- rr[no.err]
    if(length(r) < 1) stop("no non-error values")
    cdim <- unname(lapply(r, dim))
    all.null <- function(x) do.call(all, lapply(x, is.null))
    if(hasdim <- !all.null(cdim)) {
	if(length(cdim <- unique(cdim)) != 1)
	    stop("\"value\" elements of 'x' have different dimensions")
	cdim <- cdim[[1]]
	cNm <- "dimnames"
	cdmn <- lapply(r, dimnames)
    }
    else { ## not used in usual cases
	if(any(diff(clen <- vapply(r, length, 1L)) != 0))
	    stop("\"value\" elements of 'x' differ in length")
	cdim <- clen[1]
	cNm <- "names"
	cdmn <- lapply(r, names)
    }
    cdmn <- if(!all.null(cdmn)) {
	if(length(cdmn <- unique(cdmn)) != 1)
	    stop(gettextf("\"value\" elements of 'x' have different %s", cNm),
		 domain=NA)
	## if(length(cdim) > 1) cdmn[[1]] else cdmn
        ## Hmm, better if(..) ?
	if(is.list(cc <- cdmn[[1]])) cc else cdmn
    }## else NULL

    hasdim <- length(cdim) > 1 || cdim > 1
    if((dl <- length(cdim) - length(cdmn)) > 0) # e.g. for unnamed *vector* value
	## add artificial dimnames
	cdmn <- setNames(c(rep.int(list(NULL), dl), cdmn),
			 paste("D", seq_len(dl), sep="."))
    NA.proto <- r[[1]]
    NA.proto[] <- err.value
    rr[!no.err] <- list(NA.proto)
    if(is.null(FUN)) {
	FUN <- ul
	if(hasdim) {
	    dm	<- c(cdim, dm)
	    dmn <- c(cdmn, dmn)
	}
    } else stopifnot(is.function(FUN))

    array(FUN(r), dim=dm, dimnames=dmn)
}

##' @title Compute Arrays Containing the Simulation Results
##' @param x array of lists with components "value", "error", "warning",
##'	   and "time" as returned by mkAL()
##' @param comp string specifying the component to pick out
##' @param FUN function to be applied after lapply() picks out 'comp' of x
##'	       and before the array is built
##' @return an array, depending on 'comp' and FUN. The default chooses
##'	    value:        array of values or err.value (in case of an error)
##'	    error:        array of logicals indicating whether there was an error
##'	    warning:      array of logicals indicating whether there was a warning
##'	    time:         array of timings as returned by doCallWE()
##' @author Marius Hofert and Martin Maechler
##' { getArray }
getArray <- function(x, comp = c("value", "error", "warning", "time"),
		     FUN = NULL, err.value = NA)
{
    comp <- match.arg(comp)
    if(comp == "value")
	return(valArray(x, err.value=err.value, FUN=FUN))
    ## else :
    dmn <- dimnames(x)
    dm <- dim(x)
    if(is.null(FUN)) {
	FUN <-
	    switch(comp,
		   error =, warning = function(x) !vapply(x, is.null, NA),
		   time = ul)
    } else stopifnot(is.function(FUN))
    array(FUN(lapply(x, `[[`, comp)), dim=dm, dimnames=dmn)
}
##' { end }


##' { array2df }
array2df <- function(x, responseName="value") {
    rk <- length(d <- dim(x))
    ## n.sim needs to be "fixed up" {if it exists at all}:
    if(getRversion() >= "3.1.0")
	as.data.frame.table(x, responseName=responseName,
			    base = list(as.character(seq_len(prod(d[-rk])))))
    else {
	dd <- as.data.frame.table(x, responseName=responseName)
	if("n.sim" %in% names(dd))
	    dd$n.sim <- gl(d[rk], prod(d[-rk]))
	dd
    }
}
##' { end }

if(getRversion() <= "3.0.1") ##
ftable.matrix <- ftable.array <- function(x, ...) ftable(as.table(x), ...)

\end{lstlisting}
For creating the array, \code{saveSim()} calls \code{mkAL()} which is
implemented as follows:
\lstinputlisting[style=input, style=Rstyle, linerange=mkAL-end, firstnumber=1]{simsalapar/R/array-stuff.R}

%% \paragraph{maybeRead()}
%%            ===========
For reading a saved object of a simulation study, \pkg{simsalapar} provides the
function \code{maybeRead()}. If the provided \cmd{.rds} file exists,
\code{maybeRead()} reads and returns the object. Otherwise, \code{maybeRead()}
does nothing (hence the name). This is useful for reading and analyzing the result object at a
later stage by executing the same \R\ script containing both the simulation and
its analysis\footnote{Note that the first part of this paper is itself such an example.}.
\lstinputlisting[style=input, style=Rstyle, linerange=maybeRead-end, firstnumber=1]{simsalapar/R/array-stuff.R}

\subsection{Select functions for the analysis}\label{sec:select:ana}
\subsubsection[The function getArray()]{The function \code{getArray()}}
As promised in Section~\ref{sec:analysis}, we now present the implementation of
the function \code{getArray()}. This function receives the result array of lists,
picks out a specific component of the lists, and returns an array containing
these components. This is especially useful when analyzing the results of a simulation.
\lstinputlisting[style=input, style=Rstyle, linerange=getArray-end, firstnumber=1]{simsalapar/R/array-stuff.R}% getArray()

\subsubsection[The method toLatex.ftable and related functions]{The method
  \code{toLatex.ftable} and related functions}
The \code{ftable} method \code{toLatex.ftable} for creating \LaTeX\ tables calls several auxiliary
functions, detailed below.

First, the function \code{ftable2latex()} is called. It takes the provided flat
contingency table, converts \R\ expressions in the column and row variables to
\LaTeX\ expressions, and, unless they are \LaTeX\ math expressions, escapes
them (per default with the function \code{escapeLatex()}). Furthermore,
\code{ftable2latex()} takes the table entries and converts \R\ expressions (and
only those) to \LaTeX\ expressions (which are escaped in case
\code{x.escape=TRUE}; this is not the default).
\lstinputlisting[style=input, style=Rstyle, linerange=ftable2latex-end, firstnumber=1]{simsalapar/R/tables.R}

The second function called, \code{fftable()}, formats the resulting flat
contingency table (applying a new version of \code{format.ftable()} which is
available in base \R\ from 3.0.0) and returns a flat contingency table with two
attributes \code{ncv}, \code{nrv} indicating the number of column variables and
the number of row variables, respectively.

Next, \code{tablines()} is called. It receives a character matrix with
attributes \code{ncv}, \code{nrv} (typically) obtained from
\code{fftable()}. It then creates and returns a list with the
components \code{body}, \code{body.raw}, \code{head}, \code{head.raw},
\code{align}, and \code{rsepcol}. By default, \code{body} is a vector of
character strings containing the full rows (including row descriptions, if
available) of the body of the table, table entries (separated by the column
separator \code{csep}), and the row separator as specified by
\code{rsep}. \code{body.raw} provides the row descriptions (if available) and
the table entries as a character matrix. Similar for \code{head.raw} which is
a character matrix containing the entries of the table header (the number of
rows of this matrix is essentially determined by \code{ncv}); typically, this is
the header of the flat contingency table created by \code{fftable()}.
\code{head} contains a ``collapsed'' version of \code{head.raw} but in a much
more sophisticated way. \LL{\multicolumn} statements for centering of column
headings and title rules for separating groups of columns are introduced (\LL{\cmidrule} if \code{booktabs=TRUE}; otherwise
\LL{\cline}). The list component
\code{align} is a string which contains the alignment of the table entries (as
accepted by \LaTeX's \code{tabular} environment). The default implies that all
columns containing row names are left-aligned and all other columns are
right-aligned. The component \code{rsepcol} is a vector of characters which
contain the row separators \code{rsep} or, additionally, \LL{\addlinespace}
commands for separating blocks of rows belonging to the same row variables or
groups of such. The default chooses a larger space between groups of variables
which appear in a smaller column number. In other words, the ``largest'' group
is determined by the variables which appear in the first column, the
second-largest by those in the second column etc.\ up to the second-last column
containing row variables. For more details we refer to the source code of
\code{tablines()} in \pkg{simsalapar}.
%
% \lstinputlisting[style=input, style=Rstyle, linerange=tablines-end, firstnumber=1]{simsalapar/R/tables.R}

Finally, the method \code{toLatex.ftable} calls \code{wrapLaTable()}. This
function wraps a \LaTeX\ \code{table} and \code{tabular} environment around,
which can be put in a \LaTeX\ document.
\lstinputlisting[style=input, style=Rstyle, linerange=toLatex.ftable-end, firstnumber=1]{simsalapar/R/tables.R}

\subsubsection[Function mayplot() to visualize a 5D array]{%
  Function \code{mayplot()} to visualize a 5D array}
We will now present a bit more details about the function \code{mayplot()} for
creating matrix-like plots of arrays up to
dimension five. Due to space limitations, we only describe \code{mayplot()} verbally here
and refer to the source code of \pkg{simsalapar} for the exact implementation.

\code{mayplot()} utilizes the function \code{grid.layout()} to determine the
matrix-like layout, including spaces for labels; call \code{mayplot()} with
\code{show.layout=TRUE} to see how the layout looks like. \code{pushViewport()}
is then used to put the focus on a particular cell of the plot matrix (or
several cells simultaneously, see the global y axis label, for example). The
focus is released via \code{popViewport()}. Within a particular cell of the plot
matrix a panel function is chosen for plotting. This is
achieved by \pkg{gridBase}. The default panel function is either \code{boxplot.matrix()} or
\code{lines()} depending on whether \code{n.sim} exists. We also display a
background with grid lines similar to the style of \pkg{ggplot2}. Axes (for the
y axis in logarithmic scale using \code{eaxis} from \pkg{sfsmisc}) are then
printed depending on which cell the focus is on; similar for the row and
column labels of the cells, again in \pkg{ggplot2}-style. Due to the flexibility
of \pkg{grid}, we can also create a legend in
the same way as in the plot. Finally, we save initial graphical parameters with \code{opar <-
  par(no.readonly=TRUE)} and restore them on function exit in order to not
change graphical parameters for possible subsequent plots.
%
% \lstinputlisting[style=input, style=Rstyle, linerange=mayplot-end, firstnumber=1]{simsalapar/R/graphics.R}

Overall, \code{mayplot()} is quite flexible in visualizing results contained in
arrays of dimensions up to five, see the corresponding help file for more
customizations.

\subsection{Alternative varlists and simulations}\label{sec:behind:varlists}
In addition to the basic example in Section~\ref{sec:lapply}, we now call
\code{doLapply()} under various other setups, seeding methods, etc.,
including the case of \emph{no} replications, that is, \code{n.sim = 1}:

\begin{Schunk}
\begin{Sinput}
> ## doLapply() with seed=NULL (not comparable between do<parallel> methods)
> res0. <- doLapply(varList, seed=NULL, sfile="res0_lapply_NULL.rds",
                    doOne=doOne)
> ## doLapply() with seed="seq" (default)
> raw0 <- doLapply(varList, sfile="raw0_lapply_NULL.rds",
                   doAL=FALSE, ## do not call mkAL() --> keep "raw" result
                   doOne=doOne, names=TRUE)
> ## n.sim = 1 --- should also work everywhere in plot *and* table
> varList.1 <- set.n.sim(varList, 1)
> res01 <- doLapply(varList.1, sfile="res01_lapply_seq.rds", doOne=doOne,
                    names=TRUE)
> ## n.sim = 2 --- check l'Ecuyer seeding
> varList.2 <- set.n.sim(varList, 2)
> LE.seed <- c(2, 11, 15, 27, 21, 26) # define seed for l'Ecuyer
> old.seed <- .Random.seed # save .Random.seed
> set.seed(LE.seed, kind = "L'Ecuyer-CMRG") # set seed and rng kind
> (n.sim <- get.n.sim(varList.2))
\end{Sinput}
\begin{Soutput}
[1] 2
\end{Soutput}
\begin{Sinput}
> seedList <- LEseeds(n.sim) # create seed list (for reproducibility)
> system.time(
  res02 <- doLapply(varList.2, seed=seedList, sfile="res02_lapply_LEc.rds",
                    doOne=doOne, names=TRUE, monitor=interactive())      )
\end{Sinput}
\begin{Soutput}
   user  system elapsed 
  0.002   0.000   0.004 
\end{Soutput}
\begin{Sinput}
> RNGkind() # => L'Ecuyer-CMRG
\end{Sinput}
\begin{Soutput}
[1] "L'Ecuyer-CMRG" "Inversion"    
\end{Soutput}
\begin{Sinput}
> old.seed -> .Random.seed # restore .Random.seed
> RNGkind() # back to default: Mersenne-Twister
\end{Sinput}
\begin{Soutput}
[1] "Mersenne-Twister" "Inversion"       
\end{Soutput}
\end{Schunk}

%%% Local Variables:
%%% TeX-master: "parallel.tex"
%%% End:

% patchDVI setup
\Sconcordance{concordance:05a_foreach.tex:05a_foreach.Rnw:%
1 7 1 1 6 13 1 1 4 4 1 1 4 6 0 1 8 1 1 1 5 10 0 1 3 5 0 1 2 1 %
1 1 3 2 0 3 1 1 3 8 0 1 1 3 0 1 2 4 1 1 15 5 0 1 2 3 1}

% Sweave options
% use pdfcrop to crop .pdf
% put figures in ./ and name them with prefix "fig-"
%\VignetteDepends{simsalapar}
%\VignetteDepends{foreach}
%\VignetteDepends{copula}

\subsection[Using foreach]{Using \code{foreach}}
The wrapper \code{doForeach()} is based on the function \code{foreach()} of the
package \pkg{foreach}. It allows to carry out parallel computations on multiple
nodes or cores. In principle, different parallel backends can be used to conduct
parallel computations with \code{foreach()}. For example, SNOW cluster types could
be specified with \code{registerDoSNOW()} from the package \pkg{doSNOW}. We use
the package \pkg{doParallel} here which provides an interface between
\pkg{foreach} and the \R\ package \pkg{parallel}. The number of nodes can be
specified via \code{cluster.spec} (defaulting to 1) and the number of cores via
\code{cores.spec} (defaulting to \pkg{parallel}'s \code{detectCores()}). For
more details, we refer to the package source code and the vignettes of
\pkg{foreach} and \pkg{doParallel}.

\lstinputlisting[style=input, style=Rstyle, linerange=doForeach-end, firstnumber=1]{simsalapar/R/doApply.R}

Let us call \code{doForeach()} for our working example, with
\code{seed=NULL}, and \code{n.sim=1}, respectively.
\begin{Schunk}
\begin{Sinput}
> ## our working example
> res1 <- doForeach(varList, sfile="res1_foreach_seq.rds",
                    doOne=doOne, names=TRUE)
\end{Sinput}
\end{Schunk}
% Redo only this one: \rm res1_foreach_seq.rds 05a_foreach.tex 06_wrapup.tex ; make
%
\begin{Schunk}
\begin{Sinput}
> ## with seed = NULL (omitting names)
> system.time(
  res1. <- doForeach(varList, seed=NULL, sfile="res1_foreach_NULL.rds",
                     doOne=doOne))
\end{Sinput}
\begin{Soutput}
   user  system elapsed 
  0.011   0.001   0.016 
\end{Soutput}
\begin{Sinput}
> ## with n.sim = 1
> res11 <- doForeach(varList.1, sfile="res11_foreach_seq.rds",
                     doOne=doOne, names=TRUE)
\end{Sinput}
\end{Schunk}

Next, we demonstrate how l'Ecuyer's random number generator can be used.
\begin{Schunk}
\begin{Sinput}
> ## L'Ecuyer seeding (for n.sim = 2)
> old.seed <- .Random.seed # save .Random.seed
> set.seed(LE.seed, kind = "L'Ecuyer-CMRG") # set seed and rng kind
> n.sim <- get.n.sim(varList.2)
> seedList <- LEseeds(n.sim) # create seed list (for reproducibility)
> system.time(
  res12 <- doForeach(varList.2, seed=seedList, sfile="res12_lapply_LEc.rds",
  		   doOne=doOne, names=TRUE, monitor=interactive()))
\end{Sinput}
\begin{Soutput}
   user  system elapsed 
  0.000   0.000   0.004 
\end{Soutput}
\begin{Sinput}
> old.seed -> .Random.seed # restore .Random.seed
\end{Sinput}
\end{Schunk}

To see that \code{doForeach()} and \code{doLapply()} lead the same result, let
us check for equality of \code{res1} with \code{res}. We also check equality of
\code{res12} with \code{res02} which shows the same for l'Ecuyer's random number
generator.
\begin{Schunk}
\begin{Sinput}
> stopifnot(doRes.equal(res1 , res),
            doRes.equal(res12, res02))
\end{Sinput}
\end{Schunk}

%%% Local Variables:
%%% TeX-master: "parallel.tex"
%%% End:

% patchDVI setup
\Sconcordance{concordance:05b_nested_foreach.tex:05b_nested_foreach.Rnw:%
1 8 1 1 6 12 1 1 107 109 0 1 2 2 1 1 4 6 0 1 12 10 0 1 3 5 0 1 %
2 1 1 1 3 2 0 3 1 1 3 8 0 1 1 3 0 1 2 4 1 1 15 5 0 1 2 4 1}

% Sweave options
% use pdfcrop to crop .pdf
% put figures in ./ and name them with prefix "fig-"
%\VignetteDepends{simsalapar}
%\VignetteDepends{doSNOW}
%\VignetteDepends{foreach}
%\VignetteDepends{copula}

\subsection[Using foreach with nested loops]{Using \code{foreach} with nested loops}
The approach we present next is similar to \code{doForeach()}. However, it uses
nested \code{foreach()} loops to iterate over the grid variables and
replications; see the vignettes of \pkg{foreach} for the technical
details. Since this is context specific, \code{doNestForeach()} is not part of
\pkg{simsalapar}. Unfortunately, it is not possible to execute statements
between different \code{foreach()} calls. This would be interesting for
efficiently computing those quantities only once which remain fixed in subsequent
\code{foreach()} loops. Note that this is also not
possible for the other methods for parallel computing and thus not a limitation
of this method alone.

\begin{Schunk}
\begin{Sinput}
> ##' @title Function for Iterating Over All Subjobs Using Nested Foreach
> ##' @param vList list of variable specifications
> ##' @param doCluster logical indicating whether the sub jobs are run on a cluster
> ##'        or rather several cores
> ##' @param spec if doCluster=TRUE : number of nodes; passed to parallel's
> ##'                                 makeCluster()
> ##'             if doCluster=FALSE: number of cores
> ##' @param type cluster type, see parallel's ?makeCluster
> ##' @param block.size size of blocks of rows in the virtual grid which are
> ##'        computed simultaneously
> ##' @param seed see subjob()
> ##' @param repFirst see subjob()
> ##' @param sfile see saveSim()
> ##' @param check see saveSim()
> ##' @param doAL see saveSim()
> ##' @param subjob. function for computing a subjob (one row of the virtual grid);
> ##'        typically subjob()
> ##' @param doOne user-supplied function for computing one row of the (physical)
> ##'        grid
> ##' @param extraPkgs character vector of packages to be made available on nodes
> ##' @param exports character vector of functions to export
> ##' @param ... additional arguments passed to subjob() (typically further
> ##'        passed on to doOne())
> ##' @return result of applying subjob() to all subjobs, converted with saveSim()
> ##' @author Marius Hofert and Martin Maechler
> doNestForeach <- function(vList, doCluster = !(missing(spec) && missing(type)),
                            spec=detectCores(), type="MPI",
                            block.size=1, seed="seq", repFirst=TRUE,
                            sfile=NULL, check=TRUE, doAL=TRUE,
                            subjob.=subjob, doOne,
                            extraPkgs=character(), exports=character(), ...)
  {
      if(!is.null(r <- maybeRead(sfile))) return(r)
      stopifnot(is.function(doOne))
      if(!(is.null(seed) || is.na(seed) || is.numeric(seed) ||
           (is.list(seed) && all(vapply(seed, is.numeric, NA))) ||
           is.character(seed) ))
  	stop(.invalid.seed.msg)
      stopifnot(require(doSNOW), require(foreach), require(doParallel))
  
      ## variables
      pGrid <- mkGrid(vList)
      ngr <- nrow(pGrid)
      ng <- get.nonGrids(vList) # => n.sim >= 1
      n.sim <- ng$n.sim
      stopifnot(1 <= block.size, block.size <= n.sim, n.sim %% block.size == 0)
  
      ## Two main cases for parallel computing
      if(!doCluster) { # multiple cores
  	## ?registerDoParallel -> Details -> Unix + multiple cores => 'fork' is used
  	stopifnot(is.numeric(spec), length(spec) == 1)
  	registerDoParallel(cores=spec) # register doParallel to be used with foreach
      }
      else { # multiple nodes
  	## One actually only needs makeCluster() when setting up a *cluster*
  	## for working on different nodes. In this case, the 'spec' argument
  	## specifies the number of nodes.
  	## The docu about registerDoParallel() might be slightly misleading...
  	cl <- makeCluster(spec, type=type) # create cluster
  	on.exit(stopCluster(cl)) # shut down cluster and execution environment
  	registerDoParallel(cl)  # register doParallel to be used with foreach
          ## Alternative using Rmpi:
          ## cl <- makeCluster(max(2, Rmpi::mpi.universe.size()), type=type)
          ## on.exit({ ## shut down cluster and execution environment
          ##     stopCluster(cl)
          ##     if(!interactive()) Rmpi::mpi.exit() ## or directly after foreach()
          ## })
      }
      if(check) cat(sprintf("getDoParWorkers(): %d\n", getDoParWorkers()))
  
      ## need all problem-specific variables here
      ## "grid" variables
      grVals <- getEl(vList, type = "grid")
      nn <-      length(n      <- grVals$n)
      nd <-      length(d      <- grVals$d)
      nfamily <- length(family <- grVals$family)
      ntau <-    length(tau    <- grVals$tau)
  
      ## "inner" variables
      inVals <- getEl(vList, type = "inner")
      alpha  <- inVals$alpha
  
      ## actual work (note, we use a different construction here)
      n.block <- n.sim/block.size
      xpObj <- c(".Random.seed", "iter", "mkTimer", exports)
      xpPkgs <- c("simsalapar", extraPkgs)
      res <- ul(foreach(j  = seq_along(tau), .packages=xpPkgs, .export=xpObj) %:%
                foreach(k  = seq_along(family),.packages=xpPkgs,.export=xpObj)%:%
                foreach(l  = seq_along(d), .packages=xpPkgs, .export=xpObj) %:%
                foreach(m  = seq_along(n), .packages=xpPkgs, .export=xpObj) %:%
                foreach(i. = seq_len(n.block), .packages=xpPkgs, .export=xpObj)
                %dopar% {
                    i <- i. + n.block *
                        ((m-1) + nn * ((l-1) + nd * ((k-1) + nfamily * (j-1))))
                    lapply(seq_len(block.size), function(k.)
                        subjob((i-1)*block.size+k., pGrid=pGrid,
                               nonGrids=ng$nonGrids, repFirst=repFirst,
                               n.sim=n.sim, seed=seed, doOne=doOne, ...)
                           )})
      ## Now, res is a list with res[[]][[]][[]][[]][[]] corresponding to
      ## (tau, family, d, n, n.sim)
      ## ==> need to unlist (exactly the correct number of times)
      res <- ul(ul(ul(ul(res))))
      ## convert result and save
      saveSim(res, vList, repFirst=repFirst, sfile, check=check, doAL=doAL)
  }
\end{Sinput}
\end{Schunk}

Let us call \code{doNestForeach()} for our working example, with
\code{seed=NULL}, and \code{n.sim=1}, respectively.
\begin{Schunk}
\begin{Sinput}
> ## our working example
> res2 <- doNestForeach(varList, sfile="res2_nested_seq.rds",
                        doOne=doOne, names=TRUE)
\end{Sinput}
\end{Schunk}
\begin{Schunk}
\begin{Sinput}
> ## with seed = NULL (omitting names)
> system.time(
  res2.  <- doNestForeach(varList, seed=NULL, sfile="res2_nested_NULL.rds",
                          doOne=doOne)                                       )
\end{Sinput}
\begin{Soutput}
   user  system elapsed 
  0.022   0.001   0.027 
\end{Soutput}
\begin{Sinput}
> ## with n.sim = 1
> res21 <- doNestForeach(varList.1, sfile="res21_nested_seq.rds",
                         doOne=doOne, names=TRUE)
\end{Sinput}
\end{Schunk}

Next, we demonstrate how l'Ecuyer's random number generator can be used.
\begin{Schunk}
\begin{Sinput}
> ## L'Ecuyer seeding (for n.sim = 2)
> old.seed <- .Random.seed # save .Random.seed
> set.seed(LE.seed, kind = "L'Ecuyer-CMRG") # set seed and rng kind
> n.sim <- get.n.sim(varList.2)
> seedList <- LEseeds(n.sim) # create seed list (for reproducibility)
> system.time(
  res22 <- doNestForeach(varList.2, seed=seedList, sfile="res22_lapply_LEc.rds",
                         doOne=doOne, names=TRUE))
\end{Sinput}
\begin{Soutput}
   user  system elapsed 
  0.005   0.000   0.008 
\end{Soutput}
\begin{Sinput}
> old.seed -> .Random.seed # restore .Random.seed
\end{Sinput}
\end{Schunk}

To see that \code{doNestForeach()} and \code{doLapply()} lead the same result,
let us check for equality of \code{res2} with \code{res}. Finally, we check
equality of \code{res22} with \code{res02} which shows the same for l'Ecuyer's
random number generator.
\begin{Schunk}
\begin{Sinput}
> stopifnot(doRes.equal(res2,  res),
            doRes.equal(res22, res02))
\end{Sinput}
\end{Schunk}

%%% Local Variables:
%%% TeX-master: "parallel.tex"
%%% End:

% patchDVI setup
\Sconcordance{concordance:05c_Rmpi.tex:05c_Rmpi.Rnw:%
1 7 1 1 6 35 1 1 4 6 0 1 12 10 0 1 2 1 0 1 3 8 0 1 1 1 3 8 0 2 %
1 5 0 1 2 4 0 1 2 1 3 2 0 3 1 1 3 8 0 1 1 3 0 1 2 4 1 1 15 5 0 %
1 2 4 1}

% Sweave options
% use pdfcrop to crop .pdf
% put figures in ./ and name them with prefix "fig-"
%\VignetteDepends{simsalapar}
%\VignetteDepends{Rmpi}
%\VignetteDepends{copula}

\subsection[Using Rmpi]{Using \code{Rmpi}}% see http://math.acadiau.ca/ACMMaC/Rmpi/index.html
The following wrapper function \code{doRmpi()} utilizes only tools from the \R\
package \pkg{Rmpi} for parallel computing on multiple nodes or cores in \R\ via
MPI.
%% Email from Rmpi maintainer (hyu@stats.uwo.ca) on 2013-06-10:
%% If you are using OpenMPI, then mpi.universe.size() will always return 1
%% unless R is launched through mpirun.
%% Yes. You can use the option nslaves to launch slaves as many as you want.
%% How those slave processes assigned to nodes/cores are  controlled by
%% OpenMPI (different MPIs have different ways of assigning slave processes
%% but most recycle available notes/cores). In your cases, you probably
%% choose nslaves=4 so that all cores are running in parallel. However,
%% setting nslaves to be higher than the available notes/codes achieves some
%% kind loading balancing. For example, nslaves = 8 essentially spreads an
%% entire job into 8 small ones instead of 4 small ones. This gives some
%% advantages if one of the original 4 small jobs runs much longer than
%% others.
%% mpi.universe.size() # => 1; the total number of CPUs available in a cluster
%% mpi.spawn.Rslaves() # spawn as many slaves as the MPI environment knows (=> 1 master, 1 slave)
%% mpi.close.Rslaves()
%% mpi.spawn.Rslaves(nslaves=17) # spawn more slaves than possible (?) (=> 1
%%                                 master, 17 slaves) => calculations are still
%%                                 only done on the max. available cores; see
%%                                 test script http://collaborate.bu.edu/linga/ParallelMCMC
%% mpi.close.Rslaves()
%% Note: spawning more slaves than available may lead to errors (MH)
With \code{load.balancing=TRUE} (the default), the load-balancing version
\code{mpi.applyLB()} is utilized (otherwise \code{mpi.apply()}) which sends the
next sub-job to a slave who just finished one.
\lstinputlisting[style=input, style=Rstyle, linerange=doRmpi-end, firstnumber=1]{simsalapar/R/doApply.R}

Similar as before, we now call \code{doRmpi()} for our working example, with
\code{seed=NULL}, and \code{n.sim=1}, respectively. We also show here, that
\code{seed=NULL} is typically non-reproducible.
\begin{Schunk}
\begin{Sinput}
> ## our working example
> res3  <- doRmpi(varList, sfile="res3_Rmpi_seq.rds",
                  doOne=doOne, names=TRUE)
\end{Sinput}
\end{Schunk}
\begin{Schunk}
\begin{Sinput}
> ## with seed = NULL (omitting names)
> system.time(
  res3. <- doRmpi(varList, seed=NULL, sfile="res3_Rmpi_NULL.rds",
                  doOne=doOne))
\end{Sinput}
\begin{Soutput}
   user  system elapsed 
  0.011   0.000   0.014 
\end{Soutput}
\begin{Sinput}
> ## shows that seed = NULL is non-reproducible here ==> warnings (2x)
> set.seed(101)
> system.time(
  res3N1 <- doRmpi(varList, seed=NULL, sfile="res3_RmpiN1_NULL.rds",
                  doOne=doOne))
\end{Sinput}
\begin{Soutput}
   user  system elapsed 
  0.010   0.000   0.013 
\end{Soutput}
\begin{Sinput}
> set.seed(101)
> system.time(
  res3N2 <- doRmpi(varList, seed=NULL, sfile="res3_RmpiN2_NULL.rds",
                  doOne=doOne))
\end{Sinput}
\begin{Soutput}
   user  system elapsed 
  0.015   0.001   0.019 
\end{Soutput}
\begin{Sinput}
> if(identical(res3N1, res3N2)) stop("identical accidentally ??")
> str(all.equal(res3N1, res3N2)) # => they differ quite a bit!
\end{Sinput}
\begin{Soutput}
 chr [1:1644] "Component 1: Component 4: Mean relative difference: 0.007520176" ...
\end{Soutput}
\begin{Sinput}
> ## with n.sim = 1
> res31 <- doRmpi(varList.1, sfile="res31_Rmpi_seq.rds", doOne=doOne, names=TRUE)
\end{Sinput}
\end{Schunk}

\begin{Schunk}
\begin{Sinput}
> ## L'Ecuyer seeding (for n.sim = 2)
> old.seed <- .Random.seed # save .Random.seed
> set.seed(LE.seed, kind = "L'Ecuyer-CMRG") # set seed and rng kind
> n.sim <- get.n.sim(varList.2)
> seedList <- LEseeds(n.sim) # create seed list (for reproducibility)
> system.time(
  res32 <- doRmpi(varList.2, seed=seedList, sfile="res32_lapply_LEc.rds",
                  doOne=doOne, names=TRUE, monitor=interactive())      )
\end{Sinput}
\begin{Soutput}
   user  system elapsed 
  0.003   0.001   0.006 
\end{Soutput}
\begin{Sinput}
> old.seed -> .Random.seed # restore .Random.seed
\end{Sinput}
\end{Schunk}

To see that \code{doRmpi()} and \code{doLapply()} lead the same result, let
us check for equality of \code{res3} with \code{res}. We also check equality of
\code{res32} with \code{res02} which shows the same for l'Ecuyer's random number
generator.
\begin{Schunk}
\begin{Sinput}
> stopifnot(doRes.equal(res3, res),
            doRes.equal(res32,res02))
\end{Sinput}
\end{Schunk}

%%% Local Variables:
%%% TeX-master: "parallel.tex"
%%% End:

% patchDVI setup
\Sconcordance{concordance:05d_parallel_mclapply.tex:05d_parallel_mclapply.Rnw:%
1 7 1 1 6 14 1 1 2 4 0 1 5 6 0 1 12 10 0 1 3 5 0 1 2 1 1 1 3 2 %
0 3 1 1 3 8 0 1 1 3 0 1 2 4 1 1 15 5 0 1 2 4 1}

% Sweave options
% use pdfcrop to crop .pdf
% put figures in ./ and name them with prefix "fig-"
%\VignetteDepends{simsalapar}
%\VignetteDepends{parallel}
%\VignetteDepends{copula}

\subsection[Using parallel with mclapply()]{Using \pkg{parallel} with \code{mclapply()}}
Our next wrapper \code{doMclapply()} is based on the function \code{mclapply()}
of the recommended \R\ package \pkg{parallel}. Although it only parallelizes
over multiple cores, it is especially interesting to use if a larger computer
cluster is not available or if such a cluster requires complicated setup procedures. Since a
cluster is not required for \code{mclapply()} and thus \code{doMclapply()} to
work, tools like MPI need not be installed on the computer at hand. As a drawback,
this method relies on forking and hence is not available on Windows (unless the
number of cores is specified as 1 and therefore calculations are not parallel anymore).
\lstinputlisting[style=input, style=Rstyle, linerange=doMclapply-end, firstnumber=1]{simsalapar/R/doApply.R}

Let us call \code{doMclapply()} for our working example, with
\code{seed=NULL}, and \code{n.sim=1}, respectively.
\begin{Schunk}
\begin{Sinput}
> options(mc.cores = detectCores())
\end{Sinput}
\end{Schunk}
\begin{Schunk}
\begin{Sinput}
> ## our working example
> res4 <- doMclapply(varList, sfile="res4_mclapply_seq.rds",
                     doOne=doOne, names=TRUE)
\end{Sinput}
\end{Schunk}
\begin{Schunk}
\begin{Sinput}
> ## with seed = NULL (omitting names)
> system.time(
  res4. <- doMclapply(varList, seed=NULL, sfile="res4_mclapply_NULL.rds",
  		    doOne=doOne))
\end{Sinput}
\begin{Soutput}
   user  system elapsed 
  0.019   0.001   0.024 
\end{Soutput}
\begin{Sinput}
> ## with n.sim = 1
> res41 <- doMclapply(varList.1, sfile="res41_mclapply_seq.rds",
  		    doOne=doOne, names=TRUE)
\end{Sinput}
\end{Schunk}

Next, we demonstrate how l'Ecuyer's random number generator can be used.
\begin{Schunk}
\begin{Sinput}
> ## L'Ecuyer seeding (for n.sim = 2)
> old.seed <- .Random.seed # save .Random.seed
> set.seed(LE.seed, kind = "L'Ecuyer-CMRG") # set seed and rng kind
> n.sim <- get.n.sim(varList.2)
> seedList <- LEseeds(n.sim) # create seed list (for reproducibility)
> system.time(
  res42 <- doMclapply(varList.2, seed=seedList, sfile="res42_lapply_LEc.rds",
                      doOne=doOne, names=TRUE, monitor=interactive())      )
\end{Sinput}
\begin{Soutput}
   user  system elapsed 
  0.001   0.000   0.004 
\end{Soutput}
\begin{Sinput}
> old.seed -> .Random.seed # restore .Random.seed
\end{Sinput}
\end{Schunk}

To see that \code{doMclapply()} and \code{doLapply()} yield the same result, let
us check for equality of \code{res4} with \code{res}. We also check equality of
\code{res42} with \code{res02} which shows the same for l'Ecuyer's random number
generator.
\begin{Schunk}
\begin{Sinput}
> stopifnot(doRes.equal(res4, res),
            doRes.equal(res42,res02))
\end{Sinput}
\end{Schunk}

%%% Local Variables:
%%% TeX-master: "parallel.tex"
%%% End:

% patchDVI setup
\Sconcordance{concordance:05e_parallel_clusterApply.tex:05e_parallel_clusterApply.Rnw:%
1 7 1 1 6 18 1 1 5 10 0 1 3 5 0 1 2 1 1 1 3 2 0 3 1 1 3 8 0 1 %
1 3 0 1 2 3 1 1 12 4 0 1 2 3 1}

% Sweave options
% use pdfcrop to crop .pdf
% put figures in ./ and name them with prefix "fig-"
%\VignetteDepends{simsalapar}
%\VignetteDepends{parallel}
%\VignetteDepends{copula}

\subsection[Using parallel with clusterApply()]{%
  Using \pkg{parallel} with \code{clusterApply()}}
The final wrapper \code{doClusterApply()} is based on the function
\code{clusterApply()} which is the workhorse of various functions
(\code{parLapply()}, \code{parSapply()}, \code{parApply()}, etc.) in the \R\
package \pkg{parallel} for parallel computations across different nodes or
cores. In our setup, this is more efficient than calling the more well-known
wrapper function \code{parLapply()}; see the vignette of \pkg{parallel}. With
\code{load.balancing=TRUE} (the default), the load-balancing version
\code{doClusterApplyLB()} is utilized.
\lstinputlisting[style=input,style=Rstyle,linerange=doClusterApply-end,firstnumber=1]{simsalapar/R/doApply.R}

Let us call \code{doClusterApply()} with \code{seed=NULL} and \code{n.sim=1},
respectively; note that we have already called it for our working example
in Section~\ref{sec:parallel}.
%% Note that we have 'our working example' in  ./03_parallel_R.Rnw  already
%%                                              ~~~~~~~~~~~~~~~~~~
\begin{Schunk}
\begin{Sinput}
> ## with seed = NULL (omitting names)
> system.time(
  res5. <- doClusterApply(varList, seed=NULL, sfile="res5_clApply_NULL.rds",
                          doOne=doOne))
\end{Sinput}
\begin{Soutput}
   user  system elapsed 
  0.011   0.000   0.013 
\end{Soutput}
\begin{Sinput}
> ## with n.sim = 1
> res51 <- doClusterApply(varList.1, sfile="res51_clApply_seq.rds",
                          doOne=doOne, names=TRUE)
\end{Sinput}
\end{Schunk}

Next, we demonstrate how l'Ecuyer's random number generator can be used.
\begin{Schunk}
\begin{Sinput}
> ## L'Ecuyer seeding (for n.sim = 2)
> old.seed <- .Random.seed # save .Random.seed
> set.seed(LE.seed, kind = "L'Ecuyer-CMRG") # set seed and rng kind
> n.sim <- get.n.sim(varList.2)
> seedList <- LEseeds(n.sim) # create seed list (for reproducibility)
> system.time(
  res52 <- doClusterApply(varList.2, seed=seedList, sfile="res52_clApply_LEc.rds",
                          doOne=doOne, names=TRUE, monitor=interactive())       )
\end{Sinput}
\begin{Soutput}
   user  system elapsed 
  0.002   0.000   0.004 
\end{Soutput}
\begin{Sinput}
> old.seed -> .Random.seed # restore .Random.seed
\end{Sinput}
\end{Schunk}

We already checked in Section~\ref{sec:parallel} that \code{doClusterApply()}
and \code{doLapply()} lead the same result, so we only have left to check equality
for l'Ecuyer's random number generator.
\begin{Schunk}
\begin{Sinput}
> stopifnot(doRes.equal(res52,res02))
\end{Sinput}
\end{Schunk}

%%% Local Variables:
%%% TeX-master: "parallel.tex"
%%% End:

\ifTimes% patchDVI setup
\Sconcordance{concordance:06_wrapup.tex:06_wrapup.Rnw:%
1 8 1 1 5 2 1 1 8 7 0 1 2 1 5 4 0 1 5 288 0 1 2 2 1 1 2 15 0 1 %
2 4 1}

% Sweave options
% use pdfcrop to crop .pdf
% put figures in ./ and name them with prefix "fig-"
%\VignetteDepends{simsalapar}
%\VignetteDepends{doSNOW}
%\VignetteDepends{foreach}
%\VignetteDepends{copula}

\ifTimes
\section{Limited comparison of different parallelization methods}
\begin{Schunk}
\begin{Sinput}
> ## Now get the result list *per node*  ["Wish": also per "run" inside node]
> Times <- sapply(list.files("times"), function(node) {
      dir <- file.path("times", node)
      tf <- list.files(dir, pattern="do.*\\.rds$")
      names(tf) <- sub("\\.rds$",'', tf)
      t(sapply(tf, function(f) readRDS(file.path(dir, f))))
  }, simplify=FALSE)
> FF <- function(M, digits=2) format(round(M, digits=digits))
> ftab <- function(T) {
      tt <- FF(T)
      names(dimnames(tt)) <- c("method", "time")# paste("d", 1:2, sep=".")
      ftable(tt)
  }
> lapply(names(Times), function(nm) {
      ft <- ftab(Times[[nm]])
      print( toLatex(ft, caption=sprintf("Times in seconds, for machine '%s'", nm)) )
  }) -> .dev.null
\end{Sinput}
\begin{table}[htbp]
  \centering
  \begin{tabular}{*{1}{l}*{5}{r}}
    \toprule
    method \textbar\ time & \multicolumn{1}{c}{user.self} & \multicolumn{1}{c}{sys.self} & \multicolumn{1}{c}{elapsed} & \multicolumn{1}{c}{user.child} & \multicolumn{1}{c}{sys.child} \\
    \midrule
    doClusterApply\_2013-06-04 & 5.71 & 8.05 & 13.99 & 0.00 & 0.01 \\
    doClusterApply\_2013-06-18 & 9.53 & 13.42 & 23.31 & 0.00 & 0.01 \\
    doClusterApply\_2013-06-21 & 6.09 & 8.71 & 15.08 & 0.00 & 0.01 \\
    doClusterApply\_2013-06-24 & 6.20 & 8.26 & 15.42 & 0.00 & 0.02 \\
    doClusterApply\_2013-06-25 & 6.34 & 9.07 & 15.76 & 0.00 & 0.01 \\
    doClusterApply\_2013-07-01 & 5.68 & 9.15 & 15.58 & 0.00 & 0.01 \\
    doClusterApply\_2013-07-02 & 5.84 & 7.81 & 13.99 & 0.00 & 0.01 \\
    doClusterApply\_2013-07-06 & 5.85 & 8.25 & 14.40 & 0.00 & 0.01 \\
    doForeach\_2013-06-04 & 1.34 & 0.28 & 3.75 & 22.82 & 6.61 \\
    doForeach\_2013-06-18 & 1.13 & 0.22 & 3.86 & 23.97 & 4.70 \\
    doForeach\_2013-06-21 & 0.98 & 0.21 & 3.24 & 21.99 & 3.68 \\
    doForeach\_2013-06-24 & 1.12 & 0.20 & 3.54 & 24.14 & 4.94 \\
    doForeach\_2013-06-25 & 0.98 & 0.25 & 3.25 & 23.55 & 4.65 \\
    doForeach\_2013-07-01 & 0.76 & 0.18 & 3.46 & 30.54 & 4.60 \\
    doForeach\_2013-07-02 & 0.71 & 0.18 & 3.50 & 30.66 & 5.00 \\
    doForeach\_2013-07-06 & 0.71 & 0.18 & 3.42 & 29.01 & 4.57 \\
    doLapply\_2013-06-18 & 21.98 & 0.13 & 22.32 & 0.00 & 0.00 \\
    doLapply\_2013-06-21 & 20.74 & 0.12 & 21.04 & 0.00 & 0.00 \\
    doLapply\_2013-06-24 & 21.96 & 0.23 & 22.39 & 0.00 & 0.00 \\
    doLapply\_2013-06-25 & 21.83 & 0.23 & 22.23 & 0.00 & 0.00 \\
    doLapply\_2013-07-01 & 20.29 & 0.40 & 20.93 & 0.00 & 0.00 \\
    doLapply\_2013-07-02 & 20.71 & 0.21 & 21.11 & 0.00 & 0.00 \\
    doLapply\_2013-07-06 & 20.82 & 0.24 & 21.25 & 0.00 & 0.00 \\
    doMclapply\_2013-06-18 & 0.46 & 13.88 & 14.70 & 23.81 & 22.83 \\
    doMclapply\_2013-06-21 & 0.43 & 12.25 & 12.99 & 22.86 & 21.05 \\
    doMclapply\_2013-06-24 & 0.43 & 7.12 & 7.86 & 23.11 & 21.38 \\
    doMclapply\_2013-06-25 & 0.45 & 7.68 & 8.42 & 22.99 & 24.36 \\
    doMclapply\_2013-07-01 & 0.38 & 7.35 & 8.56 & 39.85 & 27.80 \\
    doMclapply\_2013-07-02 & 0.35 & 7.53 & 8.21 & 39.72 & 27.39 \\
    doMclapply\_2013-07-06 & 0.36 & 7.92 & 8.59 & 39.41 & 27.15 \\
    doNestForeach\_2013-06-18 & 2.80 & 0.26 & 5.83 & 25.03 & 5.37 \\
    doNestForeach\_2013-06-21 & 2.42 & 0.32 & 5.10 & 22.04 & 4.23 \\
    doNestForeach\_2013-06-24 & 2.65 & 0.26 & 5.20 & 23.46 & 4.96 \\
    doNestForeach\_2013-06-25 & 2.63 & 0.32 & 5.18 & 22.15 & 4.50 \\
    doNestForeach\_2013-07-01 & 1.33 & 0.20 & 4.49 & 32.10 & 5.42 \\
    doNestForeach\_2013-07-02 & 1.33 & 0.20 & 4.40 & 31.52 & 5.42 \\
    doNestForeach\_2013-07-06 & 1.38 & 0.21 & 4.40 & 29.74 & 5.00 \\
    doRmpi\_2013-06-18 & 6.42 & 11.63 & 18.71 & 0.00 & 0.04 \\
    doRmpi\_2013-06-21 & 5.17 & 9.18 & 14.93 & 0.00 & 0.03 \\
    doRmpi\_2013-06-24 & 4.71 & 8.56 & 13.84 & 0.00 & 0.04 \\
    doRmpi\_2013-06-25 & 4.96 & 8.26 & 13.72 & 0.00 & 0.03 \\
    doRmpi\_2013-07-01 & 4.26 & 8.49 & 13.22 & 0.00 & 0.04 \\
    doRmpi\_2013-07-02 & 4.58 & 8.73 & 13.86 & 0.00 & 0.04 \\
    doRmpi\_2013-07-06 & 4.60 & 8.50 & 13.79 & 0.00 & 0.04 \\
    \bottomrule
  \end{tabular}
  \caption{Times in seconds, for machine 'ada-13'}
\end{table}
\begin{table}[htbp]
  \centering
  \begin{tabular}{*{1}{l}*{5}{r}}
    \toprule
    method \textbar\ time & \multicolumn{1}{c}{user.self} & \multicolumn{1}{c}{sys.self} & \multicolumn{1}{c}{elapsed} & \multicolumn{1}{c}{user.child} & \multicolumn{1}{c}{sys.child} \\
    \midrule
    doClusterApply\_2013-06-04 & 6.52 & 10.46 & 17.16 & 0.00 & 0.02 \\
    doClusterApply\_2013-06-24 & 9.74 & 16.61 & 27.25 & 0.00 & 0.01 \\
    doClusterApply\_2013-07-03 & 5.07 & 8.05 & 13.68 & 0.00 & 0.01 \\
    doClusterApply\_2013-07-04 & 6.25 & 10.39 & 16.90 & 0.00 & 0.01 \\
    doClusterApply\_2013-07-05 & 5.08 & 8.11 & 13.47 & 0.00 & 0.01 \\
    doClusterApply\_2013-07-30 & 5.50 & 8.25 & 14.14 & 0.00 & 0.01 \\
    doForeach\_2013-06-04 & 1.10 & 0.11 & 6.31 & 16.47 & 5.85 \\
    doForeach\_2013-06-24 & 1.12 & 0.13 & 7.02 & 19.83 & 1.70 \\
    doForeach\_2013-07-03 & 0.74 & 0.08 & 4.49 & 6.36 & 0.40 \\
    doForeach\_2013-07-04 & 0.70 & 0.11 & 5.88 & 21.43 & 1.30 \\
    doForeach\_2013-07-05 & 0.75 & 0.09 & 4.40 & 12.52 & 0.81 \\
    doForeach\_2013-07-30 & 0.78 & 0.09 & 4.72 & 17.21 & 1.05 \\
    doLapply\_2013-06-24 & 21.74 & 0.24 & 22.16 & 0.00 & 0.00 \\
    doLapply\_2013-07-03 & 18.42 & 0.22 & 18.76 & 0.00 & 0.00 \\
    doLapply\_2013-07-04 & 18.86 & 0.22 & 19.24 & 0.00 & 0.00 \\
    doLapply\_2013-07-05 & 18.65 & 0.17 & 18.93 & 0.00 & 0.00 \\
    doLapply\_2013-07-30 & 18.50 & 0.21 & 18.82 & 0.00 & 0.00 \\
    doMclapply\_2013-06-24 & 0.52 & 6.77 & 13.25 & 20.84 & 18.30 \\
    doMclapply\_2013-07-03 & 0.50 & 6.41 & 8.84 & 36.31 & 21.71 \\
    doMclapply\_2013-07-04 & 0.53 & 10.03 & 12.45 & 37.45 & 25.74 \\
    doMclapply\_2013-07-05 & 0.49 & 6.99 & 9.38 & 37.28 & 22.74 \\
    doMclapply\_2013-07-30 & 0.53 & 7.47 & 9.94 & 38.19 & 24.54 \\
    doNestForeach\_2013-06-24 & 2.91 & 0.19 & 9.34 & 20.56 & 1.53 \\
    doNestForeach\_2013-07-03 & 1.52 & 0.12 & 5.27 & 11.78 & 0.89 \\
    doNestForeach\_2013-07-04 & 1.53 & 0.16 & 7.42 & 20.92 & 1.37 \\
    doNestForeach\_2013-07-05 & 1.42 & 0.10 & 5.08 & 20.51 & 1.55 \\
    doNestForeach\_2013-07-30 & 1.50 & 0.13 & 5.49 & 3.28 & 0.22 \\
    doRmpi\_2013-06-24 & 7.62 & 15.22 & 23.34 & 0.01 & 0.06 \\
    doRmpi\_2013-07-03 & 4.12 & 8.28 & 12.71 & 0.02 & 0.03 \\
    doRmpi\_2013-07-04 & 6.10 & 12.16 & 18.67 & 0.01 & 0.05 \\
    doRmpi\_2013-07-05 & 4.35 & 8.49 & 13.17 & 0.01 & 0.04 \\
    doRmpi\_2013-07-30 & 4.88 & 8.11 & 13.36 & 0.02 & 0.04 \\
    \bottomrule
  \end{tabular}
  \caption{Times in seconds, for machine 'ada-6'}
\end{table}
\begin{table}[htbp]
  \centering
  \begin{tabular}{*{1}{l}*{5}{r}}
    \toprule
    method \textbar\ time & \multicolumn{1}{c}{user.self} & \multicolumn{1}{c}{sys.self} & \multicolumn{1}{c}{elapsed} & \multicolumn{1}{c}{user.child} & \multicolumn{1}{c}{sys.child} \\
    \midrule
    doClusterApply\_2013-07-23 & 3.86 & 7.14 & 11.29 & 0.00 & 0.01 \\
    doClusterApply\_2013-07-24 & 9.62 & 21.04 & 31.41 & 0.00 & 0.01 \\
    doClusterApply\_2013-07-26 & 11.10 & 25.10 & 36.83 & 0.00 & 0.01 \\
    doClusterApply\_2013-07-31 & 4.11 & 7.59 & 12.04 & 0.00 & 0.01 \\
    doClusterApply\_2013-08-08 & 4.52 & 8.15 & 12.98 & 0.00 & 0.01 \\
    doForeach\_2013-07-23 & 0.60 & 0.10 & 3.04 & 30.54 & 4.32 \\
    doForeach\_2013-07-24 & 0.73 & 0.14 & 4.92 & 26.45 & 4.10 \\
    doForeach\_2013-07-26 & 0.82 & 0.17 & 4.91 & 30.26 & 4.45 \\
    doForeach\_2013-07-31 & 0.74 & 0.12 & 2.96 & 23.88 & 4.81 \\
    doForeach\_2013-08-08 & 0.62 & 0.11 & 2.67 & 22.46 & 4.52 \\
    doLapply\_2013-07-23 & 11.42 & 0.12 & 11.59 & 0.00 & 0.00 \\
    doLapply\_2013-07-24 & 13.97 & 0.22 & 14.28 & 0.00 & 0.00 \\
    doLapply\_2013-07-26 & 16.72 & 0.46 & 17.31 & 0.00 & 0.00 \\
    doLapply\_2013-07-31 & 11.69 & 0.22 & 12.41 & 0.00 & 0.00 \\
    doLapply\_2013-08-08 & 11.51 & 0.41 & 12.21 & 0.00 & 0.00 \\
    doMclapply\_2013-07-23 & 0.36 & 4.24 & 4.77 & 34.82 & 17.49 \\
    doMclapply\_2013-07-24 & 0.39 & 4.82 & 8.85 & 38.24 & 20.73 \\
    doMclapply\_2013-07-26 & 0.45 & 4.98 & 9.34 & 38.18 & 20.93 \\
    doMclapply\_2013-07-31 & 0.49 & 5.70 & 9.54 & 81.37 & 55.15 \\
    doMclapply\_2013-08-08 & 0.52 & 5.12 & 9.42 & 88.04 & 57.86 \\
    doNestForeach\_2013-07-23 & 1.14 & 0.11 & 3.63 & 27.78 & 4.67 \\
    doNestForeach\_2013-07-24 & 1.29 & 0.21 & 5.65 & 28.58 & 4.47 \\
    doNestForeach\_2013-07-26 & 1.35 & 0.20 & 5.86 & 27.06 & 4.23 \\
    doNestForeach\_2013-07-31 & 1.33 & 0.16 & 4.07 & 27.29 & 5.05 \\
    doNestForeach\_2013-08-08 & 1.24 & 0.18 & 4.13 & 30.07 & 4.70 \\
    doRmpi\_2013-07-23 & 3.23 & 7.52 & 11.20 & 0.01 & 0.02 \\
    doRmpi\_2013-07-24 & 7.90 & 20.71 & 29.26 & 0.00 & 0.03 \\
    doRmpi\_2013-07-26 & 8.08 & 20.94 & 29.72 & 0.01 & 0.02 \\
    doRmpi\_2013-07-31 & 3.32 & 7.76 & 11.47 & 0.00 & 0.02 \\
    doRmpi\_2013-08-08 & 3.75 & 8.90 & 13.24 & 0.01 & 0.02 \\
    \bottomrule
  \end{tabular}
  \caption{Times in seconds, for machine 'ada-7'}
\end{table}
\begin{table}[htbp]
  \centering
  \begin{tabular}{*{1}{l}*{5}{r}}
    \toprule
    method \textbar\ time & \multicolumn{1}{c}{user.self} & \multicolumn{1}{c}{sys.self} & \multicolumn{1}{c}{elapsed} & \multicolumn{1}{c}{user.child} & \multicolumn{1}{c}{sys.child} \\
    \midrule
    doClusterApply\_2013-06-04 & 7.07 & 11.17 & 18.61 & 0.00 & 0.02 \\
    doClusterApply\_2013-06-13 & 7.90 & 13.10 & 23.76 & 0.00 & 0.02 \\
    doClusterApply\_2013-06-17 & 6.47 & 10.95 & 18.17 & 0.00 & 0.01 \\
    doClusterApply\_2013-06-21 & 6.62 & 11.10 & 18.03 & 0.00 & 0.01 \\
    doClusterApply\_2013-06-24 & 6.98 & 10.94 & 18.22 & 0.00 & 0.01 \\
    doClusterApply\_2013-06-25 & 6.98 & 11.70 & 19.27 & 0.00 & 0.02 \\
    doClusterApply\_2013-06-27 & 6.55 & 11.19 & 18.81 & 0.00 & 0.02 \\
    doClusterApply\_2013-07-03 & 6.05 & 10.36 & 17.42 & 0.00 & 0.01 \\
    doForeach\_2013-06-04 & 0.98 & 0.07 & 10.71 & 13.49 & 0.53 \\
    doForeach\_2013-06-13 & 1.12 & 0.07 & 8.84 & 15.51 & 0.71 \\
    doForeach\_2013-06-17 & 0.91 & 0.09 & 6.56 & 14.35 & 0.65 \\
    doForeach\_2013-06-21 & 0.94 & 0.09 & 6.74 & 14.37 & 0.81 \\
    doForeach\_2013-06-24 & 0.94 & 0.08 & 6.75 & 15.58 & 0.75 \\
    doForeach\_2013-06-25 & 0.70 & 0.09 & 7.12 & 15.96 & 0.65 \\
    doForeach\_2013-06-27 & 0.66 & 0.08 & 7.15 & 15.82 & 0.81 \\
    doForeach\_2013-07-03 & 0.71 & 0.08 & 7.65 & 16.08 & 0.84 \\
    doLapply\_2013-06-13 & 17.35 & 0.23 & 17.89 & 0.00 & 0.00 \\
    doLapply\_2013-06-17 & 17.54 & 0.07 & 17.79 & 0.00 & 0.00 \\
    doLapply\_2013-06-21 & 17.42 & 0.08 & 17.64 & 0.00 & 0.00 \\
    doLapply\_2013-06-24 & 18.00 & 0.19 & 18.36 & 0.00 & 0.00 \\
    doLapply\_2013-06-25 & 17.00 & 0.25 & 17.41 & 0.00 & 0.00 \\
    doLapply\_2013-06-27 & 16.93 & 0.44 & 17.81 & 0.00 & 0.00 \\
    doLapply\_2013-07-03 & 17.37 & 0.23 & 17.73 & 0.00 & 0.00 \\
    doMclapply\_2013-06-13 & 0.52 & 11.40 & 23.25 & 20.03 & 17.59 \\
    doMclapply\_2013-06-17 & 0.58 & 9.16 & 14.20 & 20.29 & 16.57 \\
    doMclapply\_2013-06-21 & 0.50 & 10.10 & 14.88 & 19.83 & 18.02 \\
    doMclapply\_2013-06-24 & 0.49 & 5.85 & 11.21 & 18.54 & 16.00 \\
    doMclapply\_2013-06-25 & 0.42 & 5.54 & 10.84 & 17.95 & 14.32 \\
    doMclapply\_2013-06-27 & 0.57 & 5.87 & 17.23 & 32.46 & 23.19 \\
    doMclapply\_2013-07-03 & 0.54 & 5.87 & 16.84 & 32.00 & 23.21 \\
    doNestForeach\_2013-06-13 & 2.69 & 0.13 & 10.72 & 15.01 & 0.76 \\
    doNestForeach\_2013-06-17 & 2.27 & 0.14 & 8.31 & 9.93 & 0.53 \\
    doNestForeach\_2013-06-21 & 2.29 & 0.13 & 8.32 & 14.56 & 0.83 \\
    doNestForeach\_2013-06-24 & 2.23 & 0.13 & 7.80 & 13.90 & 0.70 \\
    doNestForeach\_2013-06-25 & 1.39 & 0.15 & 7.90 & 15.43 & 0.64 \\
    doNestForeach\_2013-06-27 & 1.33 & 0.12 & 7.93 & 16.46 & 0.76 \\
    doNestForeach\_2013-07-03 & 1.30 & 0.14 & 9.57 & 16.19 & 0.80 \\
    doRmpi\_2013-06-13 & 7.26 & 14.35 & 26.46 & 0.00 & 0.03 \\
    doRmpi\_2013-06-17 & 5.19 & 10.48 & 16.41 & 0.01 & 0.05 \\
    doRmpi\_2013-06-21 & 5.35 & 11.10 & 16.73 & 0.01 & 0.05 \\
    doRmpi\_2013-06-24 & 5.52 & 11.28 & 17.25 & 0.01 & 0.04 \\
    doRmpi\_2013-06-25 & 5.45 & 11.20 & 17.22 & 0.01 & 0.05 \\
    doRmpi\_2013-06-27 & 5.44 & 10.82 & 17.11 & 0.01 & 0.05 \\
    doRmpi\_2013-07-03 & 5.35 & 10.75 & 16.51 & 0.01 & 0.05 \\
    \bottomrule
  \end{tabular}
  \caption{Times in seconds, for machine 'lynne'}
\end{table}
\begin{table}[htbp]
  \centering
  \begin{tabular}{*{1}{l}*{5}{r}}
    \toprule
    method \textbar\ time & \multicolumn{1}{c}{user.self} & \multicolumn{1}{c}{sys.self} & \multicolumn{1}{c}{elapsed} & \multicolumn{1}{c}{user.child} & \multicolumn{1}{c}{sys.child} \\
    \midrule
    doClusterApply\_2013-06-21 & 4.65 & 18.91 & 24.12 & 0.00 & 0.00 \\
    doClusterApply\_2013-06-22 & 4.74 & 18.37 & 26.30 & 0.00 & 0.00 \\
    doClusterApply\_2013-06-25 & 3.90 & 18.52 & 23.78 & 0.00 & 0.00 \\
    doClusterApply\_2013-06-27 & 4.08 & 18.00 & 23.20 & 0.00 & 0.00 \\
    doClusterApply\_2013-07-01 & 4.73 & 23.87 & 31.11 & 0.00 & 0.00 \\
    doClusterApply\_2013-07-08 & 4.29 & 17.63 & 22.31 & 0.00 & 0.00 \\
    doClusterApply\_2013-07-11 & 4.96 & 24.50 & 32.68 & 0.00 & 0.00 \\
    doForeach\_2013-06-05 & 0.97 & 0.05 & 8.00 & 19.38 & 0.68 \\
    doForeach\_2013-06-21 & 0.82 & 0.08 & 6.91 & 16.93 & 0.71 \\
    doForeach\_2013-06-22 & 0.86 & 0.09 & 8.02 & 16.45 & 0.70 \\
    doForeach\_2013-06-25 & 0.59 & 0.06 & 7.58 & 18.58 & 0.64 \\
    doForeach\_2013-06-27 & 0.62 & 0.05 & 7.66 & 18.43 & 0.62 \\
    doForeach\_2013-07-01 & 0.65 & 0.05 & 10.30 & 18.23 & 0.52 \\
    doForeach\_2013-07-08 & 0.58 & 0.06 & 7.55 & 13.11 & 0.43 \\
    doForeach\_2013-07-11 & 0.58 & 0.08 & 8.69 & 18.62 & 0.75 \\
    doLapply\_2013-06-21 & 16.09 & 0.14 & 16.29 & 0.00 & 0.00 \\
    doLapply\_2013-06-22 & 16.04 & 0.07 & 16.16 & 0.00 & 0.00 \\
    doLapply\_2013-06-25 & 19.27 & 0.47 & 19.82 & 0.00 & 0.00 \\
    doLapply\_2013-06-27 & 18.22 & 0.16 & 18.44 & 0.00 & 0.00 \\
    doLapply\_2013-07-01 & 20.36 & 0.18 & 20.72 & 0.00 & 0.00 \\
    doLapply\_2013-07-08 & 18.88 & 0.14 & 19.08 & 0.00 & 0.00 \\
    doLapply\_2013-07-11 & 18.92 & 0.34 & 19.36 & 0.00 & 0.00 \\
    doMclapply\_2013-06-21 & 0.68 & 9.08 & 18.57 & 22.39 & 15.80 \\
    doMclapply\_2013-06-22 & 0.66 & 9.78 & 16.94 & 22.27 & 15.50 \\
    doMclapply\_2013-06-25 & 0.70 & 5.41 & 18.94 & 42.85 & 21.48 \\
    doMclapply\_2013-06-27 & 0.70 & 5.46 & 19.36 & 42.74 & 21.75 \\
    doMclapply\_2013-07-01 & 0.66 & 5.26 & 18.75 & 43.11 & 20.94 \\
    doMclapply\_2013-07-08 & 0.58 & 4.69 & 18.99 & 45.59 & 20.86 \\
    doMclapply\_2013-07-11 & 0.77 & 5.87 & 20.51 & 44.93 & 23.50 \\
    doNestForeach\_2013-06-21 & 2.06 & 0.11 & 10.01 & 18.71 & 0.61 \\
    doNestForeach\_2013-06-22 & 1.99 & 0.15 & 9.08 & 16.62 & 0.59 \\
    doNestForeach\_2013-06-25 & 1.09 & 0.07 & 8.32 & 19.50 & 0.67 \\
    doNestForeach\_2013-06-27 & 1.11 & 0.10 & 8.64 & 19.46 & 0.72 \\
    doNestForeach\_2013-07-01 & 1.31 & 0.14 & 10.80 & 19.50 & 0.61 \\
    doNestForeach\_2013-07-08 & 1.04 & 0.12 & 8.71 & 14.02 & 0.43 \\
    doNestForeach\_2013-07-11 & 1.13 & 0.12 & 10.72 & 22.02 & 0.82 \\
    doRmpi\_2013-06-21 & 3.45 & 22.01 & 28.28 & 0.00 & 0.01 \\
    doRmpi\_2013-06-22 & 2.77 & 18.96 & 24.18 & 0.00 & 0.01 \\
    doRmpi\_2013-06-25 & 2.39 & 18.70 & 21.35 & 0.00 & 0.01 \\
    doRmpi\_2013-06-27 & 2.45 & 18.00 & 22.00 & 0.00 & 0.01 \\
    doRmpi\_2013-07-01 & 3.07 & 21.60 & 25.35 & 0.00 & 0.00 \\
    doRmpi\_2013-07-08 & 2.60 & 17.72 & 20.54 & 0.00 & 0.01 \\
    doRmpi\_2013-07-11 & 2.76 & 18.38 & 25.22 & 0.00 & 0.01 \\
    \bottomrule
  \end{tabular}
  \caption{Times in seconds, for machine 'nb-mm3'}
\end{table}
\begin{table}[htbp]
  \centering
  \begin{tabular}{*{1}{l}*{5}{r}}
    \toprule
    method \textbar\ time & \multicolumn{1}{c}{user.self} & \multicolumn{1}{c}{sys.self} & \multicolumn{1}{c}{elapsed} & \multicolumn{1}{c}{user.child} & \multicolumn{1}{c}{sys.child} \\
    \midrule
    doClusterApply\_2013-06-13 & 11.21 & 41.59 & 70.55 & 0.00 & 0.00 \\
    doClusterApply\_2013-06-17 & 9.76 & 34.36 & 46.64 & 0.00 & 0.01 \\
    doClusterApply\_2013-06-18 & 9.70 & 33.38 & 44.76 & 0.00 & 0.00 \\
    doClusterApply\_2013-06-21 & 10.01 & 38.07 & 51.21 & 0.00 & 0.00 \\
    doClusterApply\_2013-06-22 & 10.44 & 37.72 & 50.84 & 0.00 & 0.00 \\
    doForeach\_2013-06-13 & 2.17 & 0.18 & 27.93 & 47.06 & 1.68 \\
    doForeach\_2013-06-17 & 1.72 & 0.11 & 17.91 & 44.69 & 1.28 \\
    doForeach\_2013-06-18 & 1.71 & 0.14 & 19.00 & 44.44 & 1.01 \\
    doForeach\_2013-06-21 & 1.73 & 0.14 & 20.54 & 44.34 & 1.18 \\
    doForeach\_2013-06-22 & 1.66 & 0.18 & 17.29 & 44.55 & 1.24 \\
    doLapply\_2013-06-13 & 55.15 & 0.61 & 57.00 & 0.00 & 0.00 \\
    doLapply\_2013-06-17 & 45.44 & 0.28 & 45.85 & 0.00 & 0.00 \\
    doLapply\_2013-06-18 & 42.73 & 0.45 & 43.33 & 0.00 & 0.00 \\
    doLapply\_2013-06-21 & 42.48 & 0.12 & 42.73 & 0.00 & 0.00 \\
    doLapply\_2013-06-22 & 43.37 & 0.10 & 43.63 & 0.00 & 0.00 \\
    doMclapply\_2013-06-13 & 1.42 & 18.20 & 59.05 & 70.71 & 33.24 \\
    doMclapply\_2013-06-17 & 1.62 & 17.10 & 42.31 & 66.11 & 32.00 \\
    doMclapply\_2013-06-18 & 1.47 & 16.41 & 35.90 & 65.39 & 31.23 \\
    doMclapply\_2013-06-21 & 1.55 & 16.25 & 40.04 & 66.06 & 31.18 \\
    doMclapply\_2013-06-22 & 1.50 & 17.23 & 33.14 & 65.20 & 31.82 \\
    doNestForeach\_2013-06-13 & 4.26 & 0.19 & 27.20 & 46.48 & 1.69 \\
    doNestForeach\_2013-06-17 & 3.52 & 0.17 & 20.63 & 45.46 & 1.24 \\
    doNestForeach\_2013-06-18 & 3.91 & 0.22 & 20.51 & 44.90 & 1.24 \\
    doNestForeach\_2013-06-21 & 3.53 & 0.18 & 21.76 & 44.25 & 1.22 \\
    doNestForeach\_2013-06-22 & 3.46 & 0.15 & 19.72 & 45.89 & 1.14 \\
    doRmpi\_2013-06-13 & 7.75 & 40.17 & 65.63 & 0.00 & 0.02 \\
    doRmpi\_2013-06-17 & 6.25 & 35.49 & 45.11 & 0.00 & 0.02 \\
    doRmpi\_2013-06-18 & 7.07 & 37.52 & 49.52 & 0.00 & 0.03 \\
    doRmpi\_2013-06-21 & 6.60 & 36.04 & 47.35 & 0.00 & 0.02 \\
    doRmpi\_2013-06-22 & 6.66 & 40.09 & 50.14 & 0.00 & 0.03 \\
    \bottomrule
  \end{tabular}
  \caption{Times in seconds, for machine 'sklar'}
\end{table}\end{Schunk}
\fi % if Times
\ifJSS\else
%% In the very end, display session info:
\begin{Schunk}
\begin{Sinput}
> toLatex(sessionInfo(), locale=FALSE)
\end{Sinput}
\begin{itemize}\raggedright
  \item R version 3.0.2 beta (2013-09-16 r63937), \verb|x86_64-unknown-linux-gnu|
  \item Base packages: base, datasets, graphics, grDevices,
    methods, parallel, stats, utils
  \item Other packages: copula~0.999-7, sfsmisc~1.0-24,
    simsalapar~1.0-0
  \item Loaded via a namespace (and not attached):
    ADGofTest~0.3, colorspace~1.2-3, grid~3.0.2,
    gridBase~0.4-6, gsl~1.9-9, lattice~0.20-21, Matrix~1.0-14,
    mvtnorm~0.9-9995, pspline~1.0-16, stabledist~0.6-6,
    stats4~3.0.2, tools~3.0.2
\end{itemize}\end{Schunk}
\fi

%%% Local Variables:
%%% TeX-master: "parallel.tex"
%%% End:
\fi

\section{Conclusion}\label{sec:conclusion}
The \R\ package \pkg{simsalapar} allows one to easily set up, conduct, and
analyze large-scale simulations studies. The user of our package only has to
provide the list of input variables on which the simulation study depends (which
can be created with the function \code{varlist()}) and the function which
computes the desired statistic (or result of the study) for one combination of input variables
(termed \code{doOne()} here). The user can then choose between different functions
to conduct the simulation (sequentially via \code{doLapply()} or in parallel via
one of \code{doForeach()}, \code{doRmpi()}, \code{doMclapply()}, or
\code{doClusterApply()}), possibly involving replicates (via a variable of type
``N'' as our \code{n.sim} here). Important aspects of a simulation study such as
catching of errors and warnings, measuring run time, or dealing with seeds are
automatically taken care of and adjusted easily. Furthermore, \pkg{simsalapar} provides
various tools to analyze the results. Besides several useful auxiliary functions,
the high-level functions \code{toLatex()} and \code{mayplot()} can be used to
create sophisticated \LaTeX\ tables and matrix-like figures of the results, respectively.

In the first part of the paper (up to and including Section~\ref{sec:analysis}),
we explained and guided the user/reader through a working example end-to-end,
which highlights various of the above steps. More advanced information about
\pkg{simsalapar}, including explanations of functions under the hood, tests, and
further examples were either addressed in the second part of the paper
(Section~\ref{sec:behind}) or can be found in the package itself; see, for
example, the demos of \pkg{simsalapar}.

\subsection*{Acknowledgements}
We would like to thank Matthias Kirchner (ETH Zurich) and Dr.\ Robin Nittka (Google
Zurich) for proofreading the manuscript and our past and current
master and Ph.D.\ students for motivating this work.

% \appendix
% Appendix will have several *section*s eventually
% each with an \input{} here:
% \input{A_mplot}% mayplot()

%% "FIXME": could generate this list automatically from ./pkgs.bib :
\nocite{murrell-gridBase%
,yu-RMpi%
,RevA-foreach%
,dahl-xtable%
,tierneyRLS-snow%
,maechler-sfsmisc%
}

\bibliography{mybib,% MMMH : Marius(& Martin)'s
  pkgs}% <-- (almost) all non-standard packages from sessionInfo()
% natbib
% \printbibliography[heading=bibintoc]% not allowed by JSS
\end{document}

%%% Local Variables:
%%% TeX-master: t
%%% End: